\documentclass[12pt,preprint]{aastex}
\bibpunct[;]{(}{)}{;}{a}{}{;}

\newcommand{\vecx}{\mbox{\boldmath $x$}}
\newcommand{\vecr}{\mbox{\boldmath $r$}}
\newcommand{\vecu}{\mbox{\boldmath $u$}}
\newcommand{\vecv}{\mbox{\boldmath $v$}}

\newcommand{\deriv}[2]{\frac{d #1}{d #2}}
\newcommand{\pderiv}[2]{\frac{\partial #1}{\partial #2}}
\newcommand{\Pderiv}[2]{\frac{\partial}{\partial #2} \left(#1 \right)}
\newcommand{\Ppderiv}[3]{\left( \frac{\partial}{\partial #2} #1 \right)_{\rm #3}}
\newcommand{\ppderiv}[3]{\left( \frac{\partial #1}{\partial #2} \right)_{\rm #3}}

\newcommand{\cint}{\int_0^{2\pi}d\theta \int_{-\infty}^\infty dz \int d \vecv}
\newcommand{\cintr}{\int_0^{2\pi} r d\theta \int_{-\infty}^\infty dz \int d \vecv}
\newcommand{\cintrr}[1]{\int_0^{2\pi} #1 d\theta \int_{-\infty}^\infty dz \int d \vecv}

\begin{document}

\shorttitle{Angular Momentum Transfer in Protolunar Disk}

\title{Angular Momentum Transfer in a Protolunar Disk}
\author{Takaaki Takeda}
\affil{Department of Earth and Planetary Sciences, Tokyo Institute of Technology}
\affil{2-12-1 Ookayama, Meguro-ku, Tokyo 152-8551, Japan}
\email{ttakeda@geo.titech.ac.jp}

\and

\author{Shigeru Ida}
\affil{Department of Earth and Planetary Sciences, Tokyo Institute of Technology}
\affil{2-12-1 Ookayama, Meguro-ku, Tokyo 152-8551, Japan}
\email{ida@geo.titech.ac.jp}

\begin{abstract}
We numerically calculated angular momentum transfer processes in a dense particulate disk within Roche limit by global $N$-body simulations, up to $N=10^5$, for parameters corresponding to a protolunar disk generated by a giant impact on a proto-Earth.
In the simulations, both self-gravity and inelastic physical collisions are included.
We first formalized expressions for angular momentum transfer rate including self-gravity and calculated the transfer rate with the results of our $N$-body simulations.
Spiral structure is formed within Roche limit by self-gravity and energy dissipation of inelastic collisions, and angular momentum is effectively transfered outward.
Angular momentum transfer is dominated by both gravitational torque due to the spiral structure and particles' collective motion associated with the structure.
Since formation and evolution of the spiral structure is regulated by the disk surface density, the angular momentum transfer rate depends on surface density, but not on particle size or number, so that the time scale of evolution of a particulate disk is independent on the number of particles ($N$) that is used to represent the disk, if $N$ is large enough to represent the spiral structure.
With $N=10^5$, the detailed spiral structure is resolved while it is only poorly resolved with $N=10^3$, however, we found that calculated angular momentum transfer does not change as long as $N \gtrsim 10^3$.
\end{abstract}

\keywords{accretion disks---Moon---planets and satellites: formation---planets: rings}

\section{INTRODUCTION}

Angular momentum transfer is essential in evolution and structure formation of accretion disks, such as galactic, plotoplanetary disks, and planetary rings.
As angular momentum is transfered outward, inner material falls to the central body and outer material migrate outward \citep{lynden1974}.
In a particulate disk within Roche limit such as a planetary ring or a plotolunar disk (see below), angular momentum is transferred through mutual collisions and self-gravitational interactions.
In the present paper, we focus on angular momentum transfer in such a disk, in particular, a protolunar disk.

``Giant Impact Hypothesis'' for the origin of the Moon \citep{hartmann1975,cameron1976} assumes that a Mars-sized protoplanet collided with the early Earth, some fraction of the impactor's mantle materials were flung away to form a circumterrestrial debris disk (a protolunar disk), and the Moon accreted from the disk.
Within Roche limit, which is about three Earth radii, accretion of large bodies is inhibited by the tidal effect from the Earth, and only materials beyond the Roche limit can form the Moon (e.g., \citet{canup1995}).
Therefore outward mass transfer from the Roche limit regulates the time scale of lunar formation (\nocite{ida1997}Ida, Canup, \& Stewart 1997; hereafter referred to as ICS97), if most materials in the protolunar disk are initially confined within Roche limit.
\citeauthor{ida1997} followed the evolution of a protolunar disk by direct $N$-body simulations with $N = 1,000$-$3,000$. 
More detailed calculations were performed by \citet*[hereafter KIM00]{kokubo2000} with $N = 10,000$.
\citeauthor{ida1997} and \citeauthor{kokubo2000} found that lunar formation time scale is very short ($\sim$ a month to a year) after condensation of disk materials.
As the protolunar disk evolves, density spiral arms are quickly developed within Roche limit.
\nocite{kokubo2000}KIM00 suggested that the spiral arms regulate the angular momentum transfer.
Also, in the local $N$-body simulations of Saturn's ring, \citet*[hereafter S95]{salo1995} and \citet*[hereafter DI99]{daisaka1999} show similar wake-like structure.

When such structure develops, disk particles move collectively.
The motion depends on surface density of the disk, independent of individual particle sizes, as shown later.
If collective motion regulates angular momentum transfer, the results in $N$-body simulations would be independent of how many particles are used for simulations, as long as particle number is enough to resolve the spiral structure.
Both physical collisions and self-gravity play important roles in the formation of the structure (\citeauthor{salo1995,daisaka1999}).

Angular momentum transfer in a planetary ring has mostly been studied in a non-self gravitating system assuming spatial uniformity.
\citet*[hereafter GT78]{goldreich1978} analytically solved the Boltzmann equation and derived an angular momentum transfer rate on the analogy of molecular viscosity. 
\citet*[hereafter AT86]{araki1986} included the effect of finite size of particles.
During a physical collision, angular momentum is transferred from one particle to another by compressive waves through interior of particles.
They showed that this type of angular momentum transfer is dominant when spatial density of particles is high enough (see section 3.1).
\citet*[hereafter WT88]{wisdom1988} formulated the angular momentum transfer in a particle disk in local coordinates, taking into account physical collisions, and numerically simulated a particle disk to calculate the transfer rate.
Their results are in good agreement with those by \citeauthor{goldreich1978} and \citeauthor{araki1986}.
\citet{petit1996} also showed similar results, numerically calculating evolution of velocity ellipsoid.

When self-gravity is included, the particles are no longer distributed uniformly (\citealt{richardson1994}; \citeauthor{salo1995,daisaka1999}).
\citet*{borderies1983} included the perturbation by a satellite and investigated the dynamics near a resonance, and \citet*{borderies1985} derived formula for angular momentum transfer by self-gravity with stream line approximation for a ring.
Angular momentum transfer due to gravitational torque has been studied also in spiral galaxies or star formation.
With trailing spirals, the angular momentum is transferred outward according to the amplitude and wavenumber of the spirals \citep{lynden1972}.
The torque exerted by the spirals transfers angular momentum effectively \citep{larson1984}.
However, the theories for galaxies do not include physical collisions.

\citet{ward1978} argued the evolution of a proto-lunar disk within Roche limit.
They assumed that clump formation by gravitational instability and destruction by tidal force would be repeated on an orbital time scale.
By estimating the energy damping rate due to inelastic collisions, they predicted the time scale of disk evolution consistent with the $N$-body simulations by \citeauthor{ida1997} and \citeauthor{kokubo2000}.
\citet{lin1987} derived a similar time scale for an accreting gas disk with turbulent viscosity induced by self-gravitational instability with dimensional analysis.
The time scale is much shorter than that predicted by \citeauthor{goldreich1978} and \citeauthor{araki1986} without self-gravity.
The self-gravity would play an essential role in the evolution of a dense particle disk system.

We formalized the angular momentum transfer in a dense particle disk (in global coordinates) including both self-gravity and collisions, starting from Boltzmann equation and generalizing \citeauthor{wisdom1988}'s formula.
We performed global $N$-body simulations including both effects, with parameters of a protolunar disk, and evaluated angular momentum transfer rate with above formalism.
We used particles up to $N = 10^5$ to represent the protolunar disk.

The protolunar disk and Saturn's ring differ in disk mass relative to the host planet mass; the ratio is 0.01-0.05 for the former while $\sim 10^{-7}$ for the latter.
The radial scale of spiral structure is given by Toomre's critical wave length (\citeauthor{salo1995,daisaka1999}), which is $\sim 2\pi(M_{\rm disk}/M_{\rm c})r_{\rm disk}$ (eq. [\ref{lambdaunstable2}]), where $M_{\rm disk}$ and $r_{\rm disk}$ are mass and radius of the disk and $M_{\rm c}$ is the planet mass.
Since the protolunar disk has much larger $M_{\rm disk}/M_{\rm p}$ than Saturnian ring, scale of spiral structure is not too small compared with the disk size, so that global calculation with $N=10^4$-$10^5$ resolves detailed spiral structure.
The angular momentum transfer in local simulations with parameters of Saturn's ring is shown in a companion paper \citep*{daisaka2001}.

We found that when $N \gtrsim$ a few thousand, the gravitational torque and collective motion regulate the angular momentum transfer.
They are both determined by the surface mass density of the disk.
When $N$ is smaller, the spiral arms are less clear, and particle motion is less collective.
However, angular momentum is transfered with nearly the same rate by gravitational interactions and random motion, as long as optical depth is not too small, which corresponds to $N \gtrsim$ several hundreds, for typical parameters of a protolunar disk. 
Therefore the result of previous works by \citeauthor{ida1997} [N=1,000-3,000], and \citeauthor{kokubo2000} [N=10,000] that the Moon formation time scale after condensation of disk materials is about a month to a year is not changed in our present calculation with $N=10^5$, although the spirals are much more clearly resolved up to fine structure in our calculation.
Actually, we performed calculations with different $N$ ($N= 10^3, 3 \times 10^3, 10^4, 3\times10^4, 10^5$) and confirmed that angular momentum transfer rate is almost independent of $N$.

Note that we must be careful to apply the above results to the evolution of a very massive disk.
Since a very massive disk such as a protolunar disk evolves very rapidly, collisional heating would dominate radiative cooling, resulting in re-vaporization of particles \citep{thompson1988}.
We will discuss this problem later (also see the discussions by \citeauthor{kokubo2000} and \citealt*{kokubo2000b}).
For less massive disks such as planetary rings, we need not worry about this problem.

In section 2 and 3, we show formulation of angular momentum transfer and how to calculate it from data of $N$-body simulations.
In section 4, we explain methods and models of our numerical simulations.
In section 5, we show the results of our simulations and discuss the angular momentum transfer in the protolunar disk. We compare the results to those by the local $N$-body simulations by \citeauthor{wisdom1988} and \citet{daisaka2001}.
In section 6, we give conclusions and discuss the problem of vaporization due to collisional heating.

\section{ANGULAR MOMENTUM TRANSFER}
\subsection{Boltzmann Equation}

In this section, we formulate angular momentum transfer in a particulate disk, including both physical collisions and self-gravity, starting from the Boltzmann equation.
The angular momentum is carried by (i) particles' random motion (like molecular viscosity), or transferred by (ii) physical collisions and (iii) gravitational forces.
During a physical collision, angular momentum is transferred by compressible waves inside the particles.
When a disk is dense enough, this contribution is large (\citeauthor{araki1986}).
We present the formulation in particle image so that the transfer rates are easily evaluated by $N$-body simulations.
We extend the formulation by \citeauthor{wisdom1988} to include the effect of gravitational torque.
Borderies et al. (1985) included gravitational torque in their formulation.
Although they adopted stream line approximation in which a disk is divided into fluid ringlets, their formulation may be essentially the same as the formulation presented here.

We assume that proto Earth is spherical, neglecting the deformation.
Non-axisymmetric deformation may remain for a while after a giant impact, which produces extra angular momentum transfer.
We will discuss the effect in section 6. 
We only count orbital angular momentum and neglect the spin angular momentum of particles, since the latter is usually much smaller than the former (\citeauthor{kokubo2000}).
We thus adopt free-slip condition that tangential restitution coefficient is unity.

Number density distribution function $f(\vecx,\vecv,m)$ satisfies the Boltzmann equation, where $\vecx$,$\vecv$, and $m$ denote position, velocity and mass of particles, respectively:
\begin{equation}
	\pderiv{f}{t} + v_{\alpha} \pderiv{f}{x_\alpha}
		- \pderiv{\Phi}{x_\alpha} \pderiv{f}{v_{\alpha}}
		= \ppderiv{f}{t}{c} + \ppderiv{f}{t}{c},
						\label{BoltzmannEq}
\end{equation}
where the derivatives with suffix $(\partial X/\partial t)_c$ and $(\partial X/\partial t)_g$ mean the change of a quantity $X$ is due to mutual collisions and gravitational interactions.
$\Phi$ is the external potential from the central body (planet).
We multiply equation (\ref{BoltzmannEq}) by $m$ and integrating it over $m$.
Since $\partial \Phi/\partial x_\alpha$ does not depend on $m$, an equation for mass density $g$ is
\begin{equation}
	\pderiv{g}{t} + v_{\alpha} \pderiv{g}{x_\alpha}
		- \pderiv{\Phi}{x_\alpha} \pderiv{g}{v_{\alpha}}
		= \ppderiv{g}{t}{c} + \ppderiv{g}{t}{g},
						\label{BoltzmannEq2}
\end{equation}
where $g (\vecx,\vecv) \equiv \int m f dm$.
In this paper, we will simulate an equal-mass ($m_0$) system, then $g=m_0f$.

We integrate equation (\ref{BoltzmannEq2}) in cylindrical coordinates over $z$, $\theta$ and $\vecv$.
After some partial integrations, we obtain the equation of continuity:
\begin{equation}
	\Pderiv{2\pi \Sigma}{t} + \frac{1}{r} \Pderiv{2\pi r \Sigma u_r}{r} =
		\Ppderiv{2\pi \Sigma}{t}{c} + \Ppderiv{2\pi \Sigma}{t}{g},
						\label{precontinuity}
\end{equation}
where $\Sigma(r,t)$ and $\vecu(r,t)$ are surface mass density and averaged velocity defined by
\begin{equation}
	\Sigma (r,t) \equiv \frac{1}{2\pi} \cint g,
						\label{DefSigma}
\end{equation}
\begin{equation}
	\Sigma \vecu (r,t) \equiv \frac{1}{2\pi} \cint \vecv g.
						\label{Defu}
\end{equation}
Since collision and gravitational force do not change the location of particles discontinuously, the right hand side of equation (\ref{precontinuity}) is 0.

Similarly, equation (\ref{BoltzmannEq2}) multiplied by $v_\theta$ with integrations over $z$, $\theta$ and $\vecv$ leads to the $\theta$ component of equation of motion,
\begin{equation}
	\Pderiv{2\pi \Sigma u_\theta}{t} 
	+ \frac{1}{r^2} \Pderiv{r^2 \cint v_r v_\theta g}{r} =
		\Ppderiv{2\pi \Sigma u_\theta}{t}{c} + \Ppderiv{2\pi \Sigma u_\theta}{t}{g}.
						\label{EqMotion}
\end{equation}

Multiplying equation (\ref{EqMotion}) by $r^2$, we obtain the equation of angular momentum,
\begin{equation}
	\Pderiv{2 \pi r \Sigma r u_\theta}{t} 
	= - \pderiv{F_{\rm AM}}{r}, 		\label{AMEq2}
\end{equation}
where
\begin{equation}
	F_{\rm AM} = F_{\rm trans}+F_{\rm col}+F_{\rm grav} + \dot{M}_{\rm disk}h
						\label{DefFAM}
\end{equation}
\begin{equation}
	F_{\rm trans} \equiv \cintr v_r r v_\theta g - \dot{M}_{\rm disk}h,
						\label{DefFmotion}
\end{equation}
\begin{equation}
	F_{\rm col} \equiv - \int_{r_{\rm min}}^r dr' 
		\Ppderiv{2 \pi r' \Sigma r' u_{\theta}}{t}{c},
						\label{DefFcol}
\end{equation}
\begin{equation}
	F_{\rm grav} \equiv - \int_{r_{\rm min}}^r dr' 
		\Ppderiv{2 \pi r' \Sigma r' u_{\theta}}{t}{g}.
						\label{DefFgrav}
\end{equation}
We introduced specific angular momentum $h \equiv ru_\theta$ and advective term $\dot{M}_{\rm disk} \equiv 2\pi r \Sigma u_{\rm r}$, which expresses angular momentum carried by mean radial flow.
Since $F_{\rm AM}$ appears in equation (\ref{AMEq2}) as $\partial F_{\rm AM}/\partial r$, the minimum range of integration $r_{\rm min}$ is a free parameter.
We choose $r_{\rm min}$ as the radius of inner boundary of the disk, so that $F_{\rm col}$ and $F_{\rm grav}$ are $0$ inside of inner boundary and outside of outer boundary of the disk.
Since $\dot{M}_{\rm disk} = 2\pi r \Sigma u_r = \cintr u_r r u_\theta g$,
\begin{equation}
	F_{\rm trans} = \cintr (v_r - u_r) r (v_\theta - u_\theta) g.
						\label{transAMflow}
\end{equation}
This term is angular momentum transfer due to random motion of particles, analogous to that due to molecular viscosity (\citeauthor{goldreich1978}).
\citeauthor{wisdom1988} and \citeauthor{araki1986} called it ``local'' and ``translational'' transfer, respectively.
Since we will use the term ``local'' in different concept, we will call this flux as ``translational'' angular momentum flux and use suffix ``trans''.
Note that in local simulation, we can set $u_{r}$ to be 0 without loss of generality (\citeauthor{wisdom1988}), but $u_r$ is not generally $0$ in global simulations.
\citeauthor{goldreich1978} neglected self-gravity and the size of particles, so that they assumed both terms in the r.h.s. of equation (\ref{EqMotion}) to be 0.
In a similar sense, \citeauthor{araki1986} neglected the second term in the r.h.s of equation (\ref{EqMotion}).
Here both terms are nonzero and $F_{\rm col}$ and $F_{\rm grav}$ have nonzero values.
Combining equation (\ref{precontinuity}) and equation (\ref{AMEq2}),
\begin{equation}
	\dot{M}_{\rm disk} \pderiv{h}{r} 
		= - 2 \pi r \Sigma \pderiv{h}{t} 
		- \Pderiv{F_{\rm trans} + F_{\rm col} + F_{\rm grav}}{r}.
						\label{massflowEq1}
\end{equation}
As long as $M_{\rm disk} \ll M_{\rm p}$, $u_\theta$ is Keplerian, $u_\theta = r\Omega = \sqrt{GM_{\rm c}/r}$, and $\partial h/\partial t = 0$.
Then 
\begin{equation}
	\dot{M}_{\rm disk} = - \frac{1}{\pderiv{h}{r}} 
		\Pderiv{F_{\rm trans} + F_{\rm col} + F_{\rm grav}}{r},
						\label{massflowEqfinal}
\end{equation}
where $\partial h/\partial r = \Omega r/2$.

Angular momentum transfer is often expressed in terms of viscosity $\nu$ (e.g. \citealt{pringle1981}). 
In a viscous accretion disk,
\begin{equation}
	\dot{M}_{\rm disk} = - 3 \pi \left( \Sigma \nu + 2 r \pderiv{\Sigma \nu}{r} \right),
						\label{massflowhikakuwakusei}
\end{equation}
\begin{equation}
	F_{\rm AM} = \dot{M}_{\rm disk} h + 3 \pi r^2 \Sigma \Omega \nu.
						\label{AMflowhikakuwakusei}
\end{equation}
Comparing equations (\ref{DefFAM}) and (\ref{massflowEqfinal}) with equations (\ref{massflowhikakuwakusei}) and (\ref{AMflowhikakuwakusei}), effective viscosity $\nu$ is given by 
\begin{equation}
	3 \pi \Sigma r^2 \Omega \nu = F_{\rm trans} + F_{\rm col} + F_{\rm grav}.
						\label{nuandF}
\end{equation}
The effective viscosities corresponding to $F_{\rm trans}$, $F_{\rm col}$ and $F_{\rm grav}$ will hereafter be denoted by $\nu_{\rm trans}$, $\nu_{\rm col}$ and $\nu_{\rm grav}$, respectively.

\subsection{Bulk and Local Random Velocities}

We further split $F_{\rm trans}$ into two parts. 
In an optically thick planetary ring system, collective particle motion exists associated with wake-like structure (\citeauthor{salo1995,daisaka1999}).
Following \citeauthor{salo1995} and \citeauthor{daisaka1999},
\begin{equation}
	\vecv = \vecv^{\rm bulk} + \vecv^{\rm local},		\label{bulklocal}
\end{equation}
where ``bulk'' velocity $\vecv^{\rm bulk}$ is locally averaged velocity (which expresses collective motion) and ``local'' velocity $\vecv^{\rm local}$ is deviation from $\vecv^{\rm bulk}$.
\citeauthor{salo1995} and \citeauthor{daisaka1999} showed that $\vecv^{\rm bulk}$ is regulated by the disk surface density but not by particle sizes.
We will discuss how to separate $\vecv^{\rm bulk}$ and $\vecv^{\rm local}$ in section 2.3.
Note that $\vecv^{\rm bulk}$ is averaged over a small range only in the vicinity of the particle, while $\vecu$ is averaged over $\theta$ from 0 to $2\pi$.

Using equation (\ref{bulklocal}), the translational angular momentum flux is divided as
\begin{eqnarray}
	F_{\rm trans} &=& \cintr (v_r^{\rm bulk} + v_r^{\rm local} - u_r ) r (v_\theta^{\rm bulk} + v_\theta^{\rm local} - u_\theta) g \nonumber	\\
	&=& \cintr (v_r^{\rm bulk} - u_r) r (v_\theta^{\rm bulk} - u_\theta) g \nonumber \\
	&+& \cintr v_r^{\rm local} r v_\theta^{\rm local} g \nonumber \\
	&=& F_{\rm bulk} + F_{\rm local}.		\label{bulklocal2}
\end{eqnarray}
Since $N$-body simulations show that $\vecv^{\rm bulk}$ and $\vecv^{\rm local}$ are not correlated, we neglected cross terms of $\vecv^{\rm bulk}$ and $\vecv^{\rm local}$ as well as those of $\vecu$ and $\vecv^{\rm local}$.

\subsection{Angular Momentum Transfer in Particle Image}

In this subsection, we derive expressions of $F_{\rm local}$, $F_{\rm col}$ and $F_{\rm grav}$ in a particle disk, generalizing the procedure by \citeauthor{wisdom1988}.
Particle $i$ is represented as $\delta(\vecx - \vecx_i) \delta(\vecv-\vecv_i) \delta (m-m_i)$ in phase space $(\vecx,\vecv,m)$, where $\vecx_i$, $\vecv_i$ and $m_i$ are the position of mass, velocity and mass of particle $i$, and $\delta$ is delta function.
Mass density distribution $g(\vecx,\vecv)$ is
\begin{equation}
	g = \int dm f = \sum_i m_i \delta(\vecx - \vecx_i) \delta(\vecv-\vecv_i).
						\label{deltafunction}
\end{equation}
With equation (\ref{deltafunction}), we rewrite $F_{\rm local}$, $F_{\rm col}$ and $F_{\rm grav}$ in particle image as below.

\subsubsection{Translational Angular Momentum Flux}

Substituting equation (\ref{deltafunction}) into equation (\ref{transAMflow}), we obtain
\begin{eqnarray}
	F_{\rm trans}(r)
		&=& \cintrr{r} 
		\left( v_r - u_r(r) \right) r 
		\left( v_\theta - u_\theta(r) \right) g	\nonumber \\
	&=& \sum_i m_i (v_{ri} - u_r(r)) r (v_{\theta i} - u_\theta(r))\delta(r-r_i).
							\label{Ftransimage0}
\end{eqnarray}
It is convenient to average $F_{\rm trans}$ over finite range $[r - \frac{1}{2} r_0 , r + \frac{1}{2} r_0]$, where $r_{\rm p} \ll r_{\rm 0} \ll r$ ($r_{\rm p}$: particle physical radius).
In our numerical simulation, we adopt $r_0 = 0.02 R_{\rm Roche}$, where $R_{\rm Roche}$ is Roche limit radius defined by equation (\ref{Roche}).
To see the sensitivity of the result to the choice of $r_0$, we also performed calculation with $r_0 = 0.05 R_{\rm Roche}$ and $r_0 = 0.1 R_{\rm Roche}$.
There is no significant difference between the cases with $r_0=0.02 R_{\rm Roche}$ and $r_0 = 0.05 R_{\rm Roche}$.
However, in the case with $r_0 = 0.10 R_{\rm Roche}$, $F_{\rm trans}$ at $r=0.4 R_{\rm Roche}$ (where $r_{\rm 0}/r$ is only 1/4) differs from $F_{\rm trans}$ with $r_0 = 0.02 R_{\rm Roche}$ by a factor 2.
Thus, $r_{\rm 0}$ should be $\lesssim 0.05 R_{\rm Roche}$.
If $r_0$ is too small, fluctuation is large, so that we have adopt $r_0 = 0.02 R_{\rm Roche}$.
Note that the other fluxes are less sensitive to the choice of $r_0$.

Averaged translational angular momentum transfer is 
\begin{eqnarray}
	\overline{F_{\rm trans}}(r) 
	&=& \frac{1}{r_0} \int_{r-\frac{1}{2}r_0}^{r+\frac{1}{2}r_0} dr' 
					F_{\rm trans}(r')	\nonumber \\
	&=& \frac{1}{r_0} \sum_{r-\frac{1}{2}r_0 < r_i < r+\frac{1}{2}r_0}
		m_i (v_{ri} - u_{r i}) r_i (v_{\theta i} - u_{\theta i}),
							\label{Ftransimage}
\end{eqnarray}
where $\vecu_i$ is averaged velocity at $\vecr_i$.

We must define the averaged velocity $\vecu_i$.
In a planetary ring, deviation of $\vecu_i$ from Keplerian velocity is usually much smaller than random velocity, so that we may replace it by Kepler velocity, $(u_{ri}, u_{\theta i}) = (0, r_i \Omega(r_i))$.
Then, equation (\ref{Ftransimage}) corresponds to the local angular momentum flux in \citeauthor{wisdom1988}.
However, since in our simulation of a protolunar disk, $|u_r|$ can not be neglected, and we do not replace $\vecu_i$ by Kepler velocity.
Substituting equation (\ref{deltafunction}) into equations (\ref{DefSigma}), and (\ref{Defu}), we obtain $\Sigma(r)$ and $\vecu(r)$ in particle image. 
We average them over $r_0$.
The averaged surface density $\Sigma \equiv \overline{\Sigma}(r)$ and averaged velocity $\vecu \equiv \overline{\vecu}(r)$ are
\begin{equation}
	\Sigma \equiv \frac{1}{2\pi rr_0} \int_{r-\frac{1}{2}r_0}^{r+\frac{1}{2}r_0} dr' \cint g 
	= \frac{1}{2\pi rr_0} \sum_j m_j, 
							\label{Sigmaimage}
\end{equation}
\begin{equation}
	\vecu \equiv 
		\frac{1}{2\pi rr_0 \Sigma } \int_{r_i-\frac{1}{2}r_0}^{r_i+\frac{1}{2}r_0} dr' \cint \vecv g 
	= \frac{1}{2\pi rr_0 \Sigma} \sum_j \vecv_j m_j,
							\label{uimage}
\end{equation}
where the summations are taken over all particles in the range $[r - \frac{1}{2} r_0 , r + \frac{1}{2} r_0]$.

Similarly, $F_{\rm bulk}$ and $F_{\rm local}$ in particle image are
\begin{equation}
	\overline{F_{\rm bulk}} = \frac{1}{r_0} \sum m_i (v_{ri}^{\rm bulk}-u_{ri}) r_i (v_{\theta i}^{\rm bulk}-u_{\theta i}),
							\label{Fbulkimage}
\end{equation}
\begin{equation}
	\overline{F_{\rm local}} = \frac{1}{r_0} \sum m_i v_{ri}^{\rm local} r_i v_{\theta i}^{\rm local}.
							\label{Flocalimage}
\end{equation}
We define $\vecv^{\rm bulk}$ as
\begin{equation}
	\vecv_i^{\rm bulk} = \vecu_i + \frac{\sum_j m_j (\vecv_j - \vecu_j)}{\sum_j m_j},
							\label{vbulkimage}
\end{equation}
where summations are taken over the range $r_i-\frac{1}{2} \lambda_{\rm bulk} < r < r_i+\frac{1}{2}\lambda_{\rm bulk}$ and $\theta_i-\frac{1}{2} \lambda_{\rm bulk}/2 \pi r_i < \theta < \theta_i+\frac{1}{2} \lambda_{\rm bulk}/2 \pi r_i$.
$\vecv^{\rm bulk}$ is mean flow velocity in the region with scale $\lambda_{\rm bulk}$.
We define $n_{\rm bulk}$ as a number of particles in the summation.
$\lambda_{\rm bulk}$ must be so large that $n_{\rm bulk} \gg 1$.
However, $\lambda_{\rm bulk}$ should be sufficiently smaller than wave length of spiral pattern.
We will use $n_{\rm bulk}$ rather than $\lambda_{\rm bulk}$ as a parameter.
Local velocity is given by $\vecv^{\rm local} = \vecv - \vecv^{\rm bulk}$.
\subsubsection{Collisional Angular Momentum Transfer}

From equation (\ref{DefFcol}),
\begin{equation}
	F_{\rm col} = - \int_{r_{\rm min}}^r dr' \cintrr{r'} r' v_\theta \ppderiv{g}{t}{c}.
						\label{Fcolimage}
\end{equation}
Collision term $(\partial g/\partial t)_{\rm c}$ is the sum of the change of distribution function $g$ over all collisions in a unit time.
During a collision between particle A and B, the locations of the particles do not change, and the velocity of each particle changes as $\vecv_{\rm A} \rightarrow \vecv_{\rm A} + \Delta \vecv_{\rm A} \;,\; \vecv_{\rm B} \rightarrow \vecv_{\rm B} + \Delta \vecv_{\rm B}$ $(m_{\rm A}\Delta \vecv_{\rm A} + m_{\rm B} \Delta \vecv_{\rm B} = 0)$.
Thus, $F_{\rm col}$ is
\begin{equation}
	F_{\rm col} = - \sum_{\rm col} \int_{r_{\rm min}}^r dr' 
	\{ m_{\rm A} \delta (r' - r_{\rm A}) \Delta v_{\theta {\rm A}}
	+ m_{\rm B} \delta (r' - r_{\rm B}) \Delta v_{\theta {\rm B}} \}
	= - \sum_{\rm col} m_{\rm A} r_{\rm A} \Delta v_{\theta{\rm A}},
						\label{Fcolimage2}
\end{equation}
where we assumed $r_{\rm A} < r_{\rm B}$ without loss of generality and used $m_{\rm A}\Delta \vecv_{\rm A} + m_{\rm B} \Delta \vecv_{\rm B} = 0$.
Summation ($\displaystyle{\sum_{\rm col}}$) is done for all collisions with $r_{\rm A} < r < r_{\rm B}$ during a unit time.

We also average $F_{\rm col}$ over the region $[r-\frac{1}{2}r_0,r+\frac{1}{2}r_0]$ to be consistent with $\overline{F_{\rm local}}$ and for the summation to have enough number of collisions.
Introducing $\Delta l = - m_{\rm A} r_{\rm A} \Delta v_{\theta {\rm A}}$,
\begin{equation}
	\overline{F_{\rm col}} = \sum_{\rm col} \frac{S}{r_0} \Delta l,
						\label{averagedFcol}
\end{equation}
where summation is taken over all collisions in the region $[r-\frac{1}{2}r_0,r+\frac{1}{2}r_0]$, during a unit time, and $S$ is radial distance by which angular momentum $\Delta l$ is transferred, i.e., $r_{\rm B} - r_{\rm A}$.
Equation (\ref{averagedFcol}) corresponds to nonlocal angular momentum flux in \citeauthor{wisdom1988}.
To avoid duplicate counting, $S = r_{\rm B} - (r - \frac{1}{2} r_0)$ or $S=(r+\frac{1}{2}r_0)-r_{\rm A}$ for the collisions which straddle the boundary $r-\frac{1}{2}r_0$ or $r+\frac{1}{2}r_0$.

\subsubsection{Gravitational Angular Momentum Transfer}

Substituting equation (\ref{deltafunction}) into equation (\ref{DefFgrav}), we obtain $F_{\rm grav}$ in particle image as
\begin{eqnarray}
	F_{\rm grav}(r) 
	&=& - \int_{r_{\rm min}}^r dr' \cintrr{r'} r' v_\theta \ppderiv{g}{t}{g}
							\nonumber \\
	&=& - \sum_{r_i < r} m_i r_i \Ppderiv{v_{\theta,i}}{t}{g} 
		= - \sum_{r_i < r} N_i^{\rm grav},
							\label{Fgravimage}
\end{eqnarray}
where $N_i^{\rm grav}$ is the torque exerted on particle $i$ by mutual gravity.
Apparently, $- F_{\rm grav}$ is total torque exerted on particles inside the radius $r$.
Since the same amount of torque is exerted on particles outside $r$ as recoils, $F_{\rm grav}$ is the angular momentum flux through a cylinder of radius $r$.
We also average $F_{\rm grav}$ over the range $[r-\frac{1}{2}r_0,r+\frac{1}{2}r_0]$,
\begin{equation}
	\overline{F_{\rm grav}}(r) = - \sum N_i^{\rm grav} 
		- \sum \frac{r_i - (r-\frac{1}{2}r_0)}{r_0} N_i^{\rm grav}, 								\label{averagedFgrav}
\end{equation}
where the first summation is taken over all particles with $r_i < r-\frac{1}{2}r_0$, and the second summation is taken over all particles in the range $[r-\frac{1}{2}r_0,r+\frac{1}{2}r_0]$.

\section{ANALYTICAL ESTIMATION OF ANGULAR MOMENTUM FLUX}

\subsection{Estimation of Angular Momentum Flux without Wake-like Structure}

In the case of a non-self gravitating particle disk, $\nu_{\rm trans} \simeq \nu_{\rm local}$ ($\nu_{\rm bulk} \simeq 0$ in this case), and it is analytically derived by \citeauthor{goldreich1978}:
\begin{equation}
	\nu_{\rm trans} \simeq \frac{\sigma^2}{\Omega}(\tau + \frac{1}{\tau})^{-1},
							\label{FtransGT}
\end{equation}
where $\sigma$ is one dimensional random velocity and $\tau$ is optical depth defined by 
\begin{equation}
	\tau = \pi r_{\rm p}^2 n_{\rm s} = \frac{\pi r_{\rm p}^2 \Sigma}{\frac{4}{3}\pi\rho_{\rm p} r_{\rm p}^3} = \frac {3 \Sigma}{4 r_{\rm p} \rho_{\rm p}},
							\label{taudefinition}
\end{equation}
where $\rho_{\rm p}$ and $n_{\rm s}$ are particle's internal density and surface number density.
A simple explanation of equation (\ref{FtransGT}) is as follows.
When collision frequency $\omega_{\rm c} \gg \Omega$, radial mean free path is $\sigma/\omega_{\rm c}$.
Then $\nu_{\rm trans} \sim \sigma \cdot \sigma/\omega_{\rm c} = \sigma^2/\omega_{\rm c}$.
When collision frequency $\omega_{\rm c} \ll \Omega$, radial mean free path is truncated at radial excursion of epicycle motion, $\sigma/\Omega$.
Since only a fraction of particles actually transfers angular momentum during one epicycle, $\omega_{\rm c}/\Omega$ must be multiplied.
Then, $\nu_{\rm local} \sim \sigma(\sigma/\Omega)(\omega_{\rm c}/\Omega) \sim \sigma^2 \omega_{\rm c} / \Omega^2$.
Replacing $\omega_{\rm c}$ by $\tau \Omega$ (\citeauthor{goldreich1978}),
\begin{equation}
	\nu_{\rm trans} \sim \cases{ \displaystyle{ \frac{\sigma^2}{\Omega}\frac{1}{\tau}} & for $\tau \gg 1$ \cr \displaystyle{ \frac{\sigma^2}{\Omega} \tau,}& for $\tau \ll 1$\cr}.
                                                        \label{nuGT1978}
\end{equation}
Interpolation of equation (\ref{nuGT1978}) gives equation (\ref{FtransGT}).
\citet{greenberg1988} gives a more detailed description.

\citeauthor{araki1986} introduced the effect of particle size in their analytical estimation, and found that at high optical depth $\nu_{\rm col} > \nu_{\rm trans}$.
Their result is consistent with numerical simulations by \citeauthor{wisdom1988}.
Their result can be fitted as \citep{schmit1995}
\begin{equation}
	\nu = \nu_{\rm col} + \nu_{\rm trans} = C \frac{\sigma^2}{\Omega} \tau^\alpha,
							\label{nufitting}
\end{equation}
where $C=0.78$ and $\alpha=1.26$.

These formulae are derived with assumption that the spatial distribution of particles is uniform and distribution in velocity space is stationary Gaussian distribution.
When wake-like structure develops in a self-gravitational disk, particles are distributed no longer uniformly, and spatially nearby particles move collectively with velocity ellipsoid like rotating bar structure (\citeauthor{daisaka1999}).
With the excitation of bulk motion, $\nu_{\rm trans}$ should be much enhanced.

\subsection{Estimation of Angular Momentum Flux with Wake-like Structure}

Gravitational angular momentum transfer in a disk with spiral pattern is analytically given by \citet{lynden1972}, under WKB approximation.
The outward angular momentum flux by self-gravitational torque is (Appendix A)
\begin{equation}
	F_{\rm grav} = \frac{\pi}{2} G r^2 \lambda_r H^2 \sin i \cos^2 i.
						\label{Fgravest}
\end{equation}
where $i$, $H$, and $\lambda_r$ are pitch angle, density amplitude, and radial wave length of spirals \citep{larson1984}.
In the linear theory, the wave length is (e.g. \citealt{binney1987})
\begin{equation}
	\lambda_{\rm cr} = \frac{2\pi^2G\Sigma}{\Omega^2}.
						\label{lambdaunstable}
\end{equation}
Note that 
\begin{equation}
	\lambda_{\rm cr} = 2\pi \frac{\pi\Sigma r^2}{M_{\rm c}} r \sim 2 \pi \frac{M_{\rm disk}}{M_{\rm c}} r,
                                                \label{lambdaunstable2}
\end{equation}
which is consistent with the results of our $N$-body simulations (Fig. \ref{snapcompare}).
In our simulations, $H\sim\Sigma$ when $\tau \sim O(1)$, and $i$ is from 15 to 45 degree (section 5.2). Here we simply put $\sin i \cos^2 i \sim O(1)$.
Though derivation of equation (\ref{Fgravest}) uses the assumption that the spiral is tightly winding ($\lambda_r \ll r$), the assumption is often good approximation even for relatively loosely winding spirals with $\lambda_r \sim r/2$ (e.g. \citealt{binney1987}).
Thus, $F_{\rm grav}$ and the corresponding viscosity are
\begin{equation}
	F_{\rm grav}^{\rm est} \sim \frac{\pi^3 G^2}{\Omega^2} r^2 \Sigma^3,
						\label{Fgravorderest}
\end{equation}
\begin{equation}
	\nu_{\rm grav}^{\rm est} \sim \frac{1}{3} \frac{\pi^2 G^2}{\Omega^3} \Sigma^2.
						\label{nugravorderest}
\end{equation}

A similar formula can be derived by the results of \citet{borderies1985}, if the results of $N$-body simulations are used.
In their model, gravitational interactions between stream lines are calculated.
The stream lines are perturbed by external potential and have the form $r(r,\phi) = a\{1-e(a)\cos m[\phi + \Delta(a)]\}$,
where $a$ and $m$ are unperturbed semi-major axis and the azimuthal mode of perturbation potential.
With negative $d\Delta/da$, these streamlines form trailing density spiral.
\cite{borderies1985} showed that in the tightly winding limit, angular momentum transfer due to gravitational torque is 
\begin{equation}
        F_{\rm grav}^{\rm est} \sim \pi^{\rm 2}G\Sigma^{\rm 2}a^{\rm 3}me^{\rm 2}.
                                                       \label{Fgravborderies}
\end{equation}
$N$-body simulations show that when spiral structure develops, $v_r^{\rm bulk} \gg v_r^{\rm local}$ and $v_r \sim v_r^{\rm bulk} \sim 2\pi G\Sigma/\Omega$ (\citeauthor{salo1995,daisaka1999}; equation (\ref{vbulkQ}) below), which means $e \sim 2\pi G \Sigma / \Omega^2 a$.
With this empirical relation and $m \sim (2 \pi a /\lambda_{\rm cr}) \tan i$, equation (\ref{Fgravborderies}) reads as
\begin{equation}
        F_{\rm grav}^{\rm est} \sim 4 \pi^{\rm 3} \frac{G^{\rm 2}}{\Omega^{\rm 2}} a^2 \Sigma^{\rm 3}.
                                                       \label{Fgravborderies2}
\end{equation}
This is similar to equation (\ref{Fgravorderest}).

\section{NUMERICAL METHODS AND MODEL}

We numerically simulated the evolution of protolunar disks by $N$-body simulations, including physical collisions and self-gravity.
We performed several sets of simulations with different initial surface density distribution.
For each set, we performed 5 runs with different particle numbers.
We started the calculation from an impact-debris disk whose mass is mostly within Roche limit.
For simplicity, we considered no heat generation and vaporization here.  
We followed collisions and gravitational interactions step by step and calculated $F_{\rm trans}$, $F_{\rm col}$ and $F_{\rm grav}$, defined as equations (\ref{Ftransimage}), (\ref{averagedFcol}) and (\ref{averagedFgrav}) at different time.
We also calculated $F_{\rm bulk}$ and $F_{\rm local}$, given by equations (\ref{Fbulkimage}) and (\ref{Flocalimage}).
The sampling time was about $100^{-1}$ Kepler time, and we averaged them over 2 Kepler time.

We set material density of the proto Earth $\rho_\oplus$ as 5.5 g ${\rm cm^{-3}}$ and that of disk particles $\rho_{\rm p}$ as 3.3 g ${\rm cm^{-3}}$, which are the present Earth's and lunar bulk densities, respectively.
With these densities, Roche limit radius is
\begin{equation}
	R_{\rm Roche} = 2.456 ( \frac{\rho_{\rm p}}{\rho_\oplus} )^{-1/3} R_\oplus = 2.9 R_\oplus,
							\label{Roche}
\end{equation}
where $R_\oplus$ is radius of the Earth.
We adopt the proto Earth mass as the present Earth mass $M_\oplus = 5.97 \times 10^{27}$ g. 
In this case, Kepler time $T_{\rm K}$ at $r = R_{\rm Roche}$ is about 7 hr.

We use $T_{\rm K}$, $M_\oplus$ and $R_{\rm Roche}$ as units of time, mass and distance.
The physical radius of a particle with mass $m$ is
\begin{equation}
	r_{\rm p}=\left( \frac{m}{M_\oplus} \right)^{1/3} \left( \frac{\rho_{\rm p}}{\rho_\oplus} \right)^{-1/3} R_\oplus = \frac{1}{2.456} \left( \frac{m}{M_\oplus} \right)^{1/3} R_{\rm Roche}.
							\label{particlesize}
\end{equation}

\subsection{Integration Method}

We follow the orbits of all particles numerically integrating the equation of motion,
\begin{equation}
	\deriv{\vecv_i}{t}
	= - G M_\oplus \frac{\vecx_i}{|\vecx_i|^3} - \sum^N_{j \neq i} G m_i \frac{\vecx_i-\vecx_j}{|\vecx_i - \vecx_j|}.
							\label{Newton'slaw}
\end{equation} 
We use fourth ordered Hermite scheme with individual time step scheme.
Full description of the scheme is given in \citet{makino1992}.
Let $d$ be distance between mass centers of two particles.
When we detect $d$ smaller than the sum of particle radii $(r_{\rm p1} + r_{\rm p2})$, we change velocities according to the restitution coefficient.
For detailed adjustment of collision and rebounding to avoid numerical difficulty, we follow \citet{richardson1994}.

Since particle number is limited in N-body simulation, fragmentation can not be included in practice.
As shown below, typical random velocity of particles is $\sim \pi G \Sigma/\Omega$, which ranges from a few hundreds ${\rm ms^{-1}}$ to a few ${\rm kms^{-1}}$.
The corresponding specific energy is from $2 \times 10^8$ to $2 \times 10^{10} {\rm ergg^{-1}}$.
A collision with such energy may results in a catastrophic disruption (e.g. \citealt{benz1999}) unless particles are small enough that material strength is important.
However, the angular momentum transfer depends on surface density, but not on each particle size, as shown below.
The disruption would not affect the angular momentum transfer process.
If a disruptive collision is supposed to occur, we represent energy dissipation during the disruption as an effective restitution coefficient for inelastic collision.
Since we do not know the value of the effective restitution coefficient, we perform runs with different values.

We assume that tangential restitution coefficient $\varepsilon_{\rm t}=1$, neglecting the exchange between orbital angular momentum and spin angular momentum (KIM00).
For simplicity, we use normal restitution coefficient $\varepsilon_{\rm n}$ that is independent of collision velocity.
The mean reduction of relative velocity due to the inelasticity is \citep{canup1995}
\begin{equation}
	1 - \varepsilon = 1 - \left[  \frac{\varepsilon_{\rm n}^2 v_{\rm n}^2+\varepsilon_{\rm t}^2v_{\rm t}^2}{v_{\rm n}^2+v_{\rm t}^2} \right]^{1/2}.
\end{equation}
Assuming mean tangential and normal collision velocities, $v_{\rm t}$ and $v_{\rm n}$, are similar, effective restitution coefficient $\varepsilon$ is about 0.7, for example, if $\varepsilon_{\rm n} = 0.1$.
This corresponds to that about half ($\sim 0.7^2$) the fraction of relative kinetic energy of collision is given to fragments.
In nominal cases, we adopt $\varepsilon_{\rm n} = 0.1$, which may be rather dissipative one. 
We also study the effect of the restitution coefficient with additional set of simulations with different $\varepsilon_{\rm n}$ from 0.1 to 0.8.
We will show that angular momentum transfer rates are almost independent of $\varepsilon_{\rm n}$ except for highly elastic $\varepsilon_{\rm n}$ ($\varepsilon_{\rm n} \gtrsim 0.6$); in the highly elastic cases, velocity distribution is so high that spiral arms are not developed.

If the relative velocity after a collision is sufficiently small, the collision results in gravitationally bounded particles.
At $r \gtrsim R_{\rm Roche}$, the particles are gravitationally bounded if Jacobi energy is negative and particles are within Hill sphere \citep{ohtsuki1993,canup1995,kokubo2000b}.
We adopt rubble pile model, in which no mergers are allowed.
In this case, gravitationally bounded particles form particle aggregates.
When the aggregates are scattered to inside of Roche limit, particle size overflows Hill sphere and tidal force disrupts the aggregates \citep{ohtsuki1993,canup1995,kokubo2000b}.

\subsection{Parameters and Initial Conditions}

We simulate disks with equal-mass particles.
The initial surface density $\Sigma$ is distributed as $\Sigma \propto a^{\alpha}$ in the range $a_{\rm min}<a<a_{\rm max}$, where $a$ is semimajor axis.
The initial conditions of the disks are shown in Table I. 
We set $a_{\rm max}$ near Roche limit, so that most particles are initially within Roche limit.
SET 1-4 start from centrally condensed disks with $\alpha = - 3$, with different total disk masses.
Runs with constant surface density ($\alpha = 0$) are investigated in SET 5 in comparison.
We simulated the disks by runs with $N= 10^3, 3 \times 10^3, 10^4, 3\times10^4$ and $10^5$ for each SET.
Totally we performed 25 runs with $\varepsilon_{\rm n} = 0.1$. We call a run with $j\times1000$ particles in SET $k$ as ``RUN $k$-$j$K''.
We also performed additional simulations (SET 6), with fixed initial particle number $N=3 \times 10^4$, changing the normal restitution coefficient as $\varepsilon_{\rm n} = 0.2, 0.4, 0.6$ and $0.8$.
For each $\varepsilon_{\rm n}$, we perform 4 runs with different initial surface density distribution similar to SET 1-4 (SET6 includes 16 runs).

Orbital eccentricities and inclinations are given by Rayleigh distribution with given mean values $\langle e^2 \rangle^{1/2}$ and $\langle i^2 \rangle^{1/2}$.
Specific choice of initial $\langle e^2 \rangle^{1/2}$ and $\langle i^2 \rangle^{1/2}$ does not affect the results because they approach quasi-equilibrium values only on one or two Kepler times.

\section{RESULTS}
\subsection{Overall Disk Evolution}

We simulated the evolution of a plotolunar disk and investigated angular momentum transfer, which causes mass redistribution of the disk.
First we show overall evolution of the disk.
In Figs. \ref{snap} (a)-(f), we show snapshots of disk evolution for SET 2 with $1 \times 10^5$ particles (RUN 2-100K).
We also show snapshots in $r$-$z$ plane in Figs. \ref{snapedge} (a)-(f).
At $t=0$ (panel a), particles are distributed azimuthally uniformly.
In SET 2, we give initially high random velocity, so that the disk scale hight is large (Fig. \ref{snapedge}(a)) to avoid self-gravitational instability initially.
Inelastic collisions quickly damp the random motion.
Since the orbital rotation time is shorter and optical depth is higher in inner region, the damping is faster there.
At $t=2 T_{\rm K}$ (panel b), the spirals begin to develop in the inner region.
However, as we will discuss later, clear spiral structure does not develop in the innermost region.
At $t=6 T_{\rm K}$ (panel c) and $t=10 T_{\rm K}$ (panel d), spirals extend to the outer region, and disk material is transferred beyond the Roche limit.
At $t=20 T_{\rm K}$ (panel e), lunar seeds (large aggregates) are formed beyond the Roche limit.
At $t=30 T_{\rm K}$ (panel f), two large lunar seeds further grow.
The largest and the second largest seeds consist of about 5000 and 2400 particles respectively in this stage.
These lunar seeds grow to a single moon.
This is typical evolution of the protolunar disk (\citeauthor{ida1997,kokubo2000}).

In Figs. \ref{snapcompare} (a)-(e), we show snapshots of RUN 1-100K, 2-100K, 3-100K, 4-100K, with different $M_{\rm disk}$ from about $1.5 M_{\rm L}$ to $5 M_{\rm L}$.
The radial wave length of spirals is larger in heavier disks as predicted by equation (\ref{lambdaunstable}).
Equation (\ref{lambdaunstable}) agrees well with numerical results, as shown in section 5.2.

We show time evolution of surface density distributions of RUN 2-100K and RUN 5-30K in Figs. \ref{surden1} (a) and (b).
In RUN 5-30K surface density is initially flat.
After several Kepler times, quasi-equilibrium state is achieved in both cases, and disk mass is rather steadily transferred.
In the outer region, mass is transferred outward to form lunar seeds, which correspond to a small peak of surface density beyond Roche limit.
In the inner part, particles fall onto the Earth in compensation, and surface density decreases.

As we will show in the following sections, effective viscosity is proportional to $r_{\rm p}^{1/2}\Sigma^{3/2}$ in the region $r \lesssim 0.6 R_{\rm Roche}$ (section 5.3.1), and to $\Sigma^3$ in the region $r \gtrsim 0.6 R_{\rm Roche}$ (section 5.3.2).
Collisional angular momentum transfer is dominant at $r \lesssim 0.6 R_{\rm Roche}$, while other two processes are dominant at $r \gtrsim 0.6 R_{\rm Roche}$.
In Figs. \ref{surden1} (c) and (d), we show the surface density distribution at $t = 6 T_{\rm K}$ and $t=18 T_{\rm K}$ of SET 2 with various $N$.
In the inner region, the peak hight of surface density is lower for smaller $N$, because $\nu \propto r_{\rm p}^{1/2}\Sigma^{3/2}$ increases with decrease in $N$.
On the other hand, the surface density distribution in the outer region is almost independent of $N$, because effective viscosity is independent of physical size $r_{\rm p}$.
In this region, angular momentum transfer is regulated by collective effects, so that $\nu$ is dependent only on $\Sigma$, and does not depend on $r_{\rm p}$.  
The lunar accretion processes are regulated by mass and angular momentum transfer near $R_{\rm Roche}$ as explained in \citeauthor{ida1997} and \citeauthor{kokubo2000}.
Figs. \ref{surden1} (c) and (d) indicate they are almost independent of $r_{\rm p}$ or particle number $N$ of N-body simulation.

We show magnified snapshots of SET 2 in Figs. \ref{closeup} (a) and (b) and SET 3 in Figs. \ref{closeup} (c) and (d).
Figs. \ref{closeup} (a) and (c) are $N = 1 \times 10^5$ cases and (b) and (d) are $N = 1 \times 10^4$ cases.
Spiral structure with high density contrast develops only in the outer region.
To clarify quantitative features of spiral structure, we adopt autocorrelation analysis.
\subsection{Autocorrelation Analysis for spiral structure}

We calculated autocorrelation directly by superposing the relative locations of particles, following \citeauthor{salo1995} and \citeauthor{daisaka1999} with a little modification.
The autocorrelation of spatial distribution function $n$ is
\begin{equation}
        {\rm Corr}(\Delta r,\Delta\theta,r) = \frac{1}{f_{\rm n}} \int_{r-\frac{1}{2}r_0}^{r+\frac{1}{2}r_0} dr' \int_0^{2\pi} r'd\theta n(r'+\Delta r, \theta + \Delta\theta) n(r',\theta),
                                                        \label{correlationdef}
\end{equation}
where $\Delta r$, $\Delta \theta$ and $f_{\rm n}$ are relative locations and normalizing factor.
As shown below, ${\rm Corr}$ is a function of $r$.
Substituting $n(r,\theta) = \sum m_i \delta(r-r_i) \delta(\theta-\theta_i)/r$ to equation (\ref{correlationdef}),
\begin{eqnarray}
	{\rm Corr}(\Delta r,\Delta\theta,r) &=& \frac{1}{f_{\rm n}} \sum_i \sum_j m_i m_j \int_{r-\frac{1}{2}r_0}^{r+\frac{1}{2}r_0} dr' \frac{\delta(r'+\Delta r - r_i )}{r'+\Delta r} \delta(r'-r_j) \nonumber \\
        && \times \int_{r-\frac{1}{2}r_0}^{r+\frac{1}{2}r_0} d\theta \delta(\theta+\Delta \theta - \theta_i) \delta(\theta - \theta_j) \nonumber \\
         &=& \frac{1}{f_{\rm n}} \sum_i \sum_j \frac{m_i m_j}{r_i} \delta(\Delta r-(r_i-r_j)) \delta(\Delta \theta - (\theta_i - \theta_j)),
                                                        \label{correlation}
\end{eqnarray}
where $j$ is summed over particles in the region $[r-\frac{1}{2}r_0, r+\frac{1}{2}r_0]$ for each reference radius $r$, while summation over $i$ is taken for all particles.
Equation (\ref{correlation}) is superposition of particles' relative distribution.
We used $60\times200$ mesh in $(\Delta r, \Delta \theta)$ space in the region $\Delta r =[-0.3, 0.3]$, and $\Delta \theta = [-1, 1]$ (radian) and averaged ${\rm Corr}$ in each mesh.
Since radial density gradient is large, we choose following normalizing factor,
\begin{equation}
	f_{\rm n} = 2\pi \int_{r-\frac{1}{2}r_0}^{r+\frac{1}{2}r_0} dr' r' \Sigma(r' + \Delta r)\Sigma(r').
\end{equation}
With above normalizing factor, ${\rm Corr}$ is 1 everywhere if surface density is a function only of $r$.

Since unstable wave length depends on $\Sigma$, ${\rm Corr}$ depends on surface density.
In global simulation, $\Sigma$ changes with time, so that we choose snapshot data from different runs with similar surface density and averaged them.
We choose only snapshot with initial $N \geq 3 \times 10^4$ and $\varepsilon_{\rm n} \leq 0.2$.
We show the contour maps of ${\rm Corr}$ in Figs. \ref{figcorrelation}.
We averaged 10 to 30 ${\rm Corr}$ snapshots data to draw each figure.
The straight dotted lines represent pitch angles with $15, 30$ and $45$ degrees.
We show ${\rm Corr}$ at $r=0.7 R_{\rm Roche}$ with different surface density in panel (a) $\Sigma = 0.005$-$0.006$, (b) $\Sigma = 0.008$-$0.010$ and (c) $\Sigma = 0.012$-$0.016$.
Pitch angle is always $\sim 20$ degrees and it does not depend on surface density within the range of our simulation.
However, we found that pitch angle increases with $r$.
Similar tendency is also shown in local simulations (\citeauthor{salo1995} and \citeauthor{daisaka1999}).
Panel (d), (e), and (f) show ${\rm Corr}$ at $r=0.5 R_{\rm Roche}, 0.6 R_{\rm Roche}$ and $0.9 R_{\rm Roche}$, with $\Sigma =$ 0.012-0.016, 0.012-0.016, and 0.004-0.005.
In innermost region (d), there is no clear structure as shown in Figs. \ref{closeup}, and ${\rm Corr}$ is almost featureless.
In outer region ((e) and (f)), clear bar like structure appears at the center.
The pitch angle is about 15 degree at $r=0.6 R_{\rm Roche}$ and about 45 degree at $r=0.9 R_{\rm Roche}$.
In the analytical estimate in the following sections, we adopt 30 degrees as an averaged pitch angle.

In panel (a) or (b), we also recognize the bars next to the central bars.
The radial separation increases with $\Sigma$.
The corresponding $\lambda_{\rm cr}$ (eq. [\ref{lambdaunstable}]) for panel (a), (b) and (c) are 0.046$R_{\rm Roche}$, 0.061$R_{\rm Roche}$ and 0.095$R_{\rm Roche}$.
Our numerical result and linear theory agree well.

\subsection{Angular Momentum Transfer}

Figs.\ref{AM} (a), (b), (c) and (d) show angular momentum fluxes $F_{\rm col}$, $F_{\rm grav}$ and $F_{\rm trans}$ in RUN 2-100K, RUN2-10K, RUN 3-100K and RUN 3-10K, after spiral patterns fully develop but before large lunar seeds grow, which may perturb the entire disk.
The horizontal axis is distance from the center of the Earth, and vertical axis is angular momentum flux.
In inner region ($r \lesssim 0.6 R_{\rm Roche}$), collisional angular momentum flux $F_{\rm col}$ dominates.
In outer region ($0.6 R_{\rm Roche} \lesssim r$), $F_{\rm grav}\sim F_{\rm trans} > F_{\rm col}$.
The difference should reflect the fact that spiral structure develops in the outer region while it does not in the inner region.

Before discussing angular momentum fluxes in each region in detail, we consider the condition for development of spiral structure. 
In the linear theory, axisymmetric density perturbations grow if $Q < 1$ where $Q$ is defined by $\sigma \Omega/ \pi G \Sigma$ \citep{toomre1964} with velocity dispersion $\sigma$.
We show $Q$ values as a function of $r$ averaged over from $t=8 T_{\rm K}$ to $t=10 T_{\rm K}$ for SET 2, and from $t=4 T_{\rm K}$ to $t=6 T_{\rm K}$ for SET 3 in Figs.\ref{Qvalue}. 
In outer region, velocity of collective motion is pumped up by spirals and $Q$ is greater than 1.
Local simulations show that $Q \sim 2$ when spirals develop (\citeauthor{salo1995,daisaka1999}).
In our simulation, $Q$ increases with $r$, up to about 5 at $r = R_{\rm Roche}$, which may be caused by global effects.
In the inner region where optical depth is high, $Q$ is enhanced by incompressibility of the particles.
Physical meaning of $Q$ is $\sim \Omega^2/\pi G \rho$, where spatial density $\rho \sim \Sigma/h$ and disk scale hight $h \sim \sigma/\Omega$ are used.
In the high optical depth case, $\rho$ is limited by particle material density $\rho_{\rm p}$ and $Q$ should be $\gtrsim \Omega^2/\pi G \rho_{\rm p}$.
The minimum value of $Q$ is independent of $\sigma$ and $r_{\rm p}$ (or $N$), and increases with decrease in $r$.
\citet{ward1978} argued that mid-plane pressure stabilizes the disk and predicted that gravitational instability does not occur when $r \lesssim 0.5 R_{\rm Roche}$ even if random velocity is small.
A similar criterion is also derived by comparing particle's Hill radius with physical size (\citeauthor{daisaka1999}).
Our result that spiral structure does not develop in the inner region is consistent with these predictions.

\subsubsection{Angular Momentum Transfer in Inner Region}

In Fig. \ref{nucoltau}, we plot $\nu_{\rm col}/r_{\rm p}^2\Omega$ obtained by our simulations at $r=0.5 R_{\rm Roche}$, at various times of all RUNs with various $\Sigma$ and $r_{\rm p}$.
Collisional viscosity $\nu_{\rm col}$ is well fitted as
\begin{equation}
	\nu_{\rm col} \simeq 1.35 \Omega r_p^2 \tau^{1.5}.
							\label{nucolnumerical}
\end{equation}
We found $\nu_{\rm col}$ is consistent with the results of \citeauthor{araki1986}.
In the inner region, one dimensional velocity dispersion $\sigma \sim r_{\rm p} \Omega$ (\citeauthor{salo1995}).
Thus, equation (\ref{nufitting}) reads as
\begin{equation}
	\nu = \nu_{\rm col} + \nu_{\rm local} \sim 0.78 \Omega r_{\rm p}^2 \tau^{5/4}.
							\label{estSM}
\end{equation}
In Fig. \ref{nucoltau}, we also plot $(\nu - \nu_{\rm local})/r_{\rm p}^2 \Omega$ (solid line), where we use equation (\ref{FtransGT}) for $\nu_{\rm local}$.
The results of our N-body simulation agree with equation (\ref{estSM}) except for a numerical factor $\sim$ $2$-$3$.
$\nu_{\rm col}$ also agrees with the result of local simulations when the spiral structure does not develop \citep{daisaka2001}.

Since $\tau \propto \Sigma r_{\rm p}^{-1}$ (eq.[\ref{taudefinition}]), $\nu_{\rm col} \propto r_{\rm p}^{1/2} \Sigma^{3/2}$.
For the same $\Sigma$, $\nu_{\rm col}$ decreases with decrease in $r_{\rm p}$, that is, increase in $N$.
From equations (\ref{nuandF}) and (\ref{nucolnumerical}),
\begin{equation}
	F_{\rm col} = 3\pi \Sigma r^2 \Omega \nu_{\rm col} \simeq 8.26 r^2 \Omega^2 \Sigma^{5/2} \rho^{-3/2} r_{\rm p}^{1/2}.				\label{Fcol05}
\end{equation}
Eliminating $r_{\rm p}$ using equation (\ref{taudefinition}), $F_{\rm col}$ at $r=0.5$ is
\begin{equation}
	F_{\rm col} \simeq C_{\rm c} \tau^{-1/2} \left( \frac{\pi^3 G^2}{\Omega^2} r^2 \Sigma^3 \right),
							\label{Fcol05toCc}
\end{equation}
where $C_{\rm c} \simeq 1.2$.

\subsubsection{Angular Momentum Transfer In Outer Region}

\subsubsubsection{5.3.2.1 Angular Momentum Transfer by Gravitational Torque}
In the outer region, where spiral structure develops, $F_{\rm grav}$ and $F_{\rm trans}$ overwhelm $F_{\rm col}$.
We show $F_{\rm grav}$ as a function of surface density at $r=0.7 R_{\rm Roche}$ in Fig. \ref{Fgrav} (a).
We plot the results of all RUNs in SET 1-4, during $t=6 T_{\rm K}$ to $t=16 T_{\rm K}$ with sampling intervals $2 T_{\rm K}$.
Since $\Sigma$ is different between different RUNs and between different times, we obtain $F_{\rm grav}$ with various $\Sigma$ at the same $r$.
The analytical estimation of $F_{\rm grav}^{\rm est} = (\pi^3 G^2 / \Omega^2) r^2 \Omega^3 \sin i \cos^2 i$ (eq.[\ref{Fgravorderest}]) with $i=30$ degrees is represented by a dashed line.
Angular momentum flux $F_{\rm grav}$ in numerical results is in agreement with $F_{\rm grav}^{\rm est}$, in particular in the case of high $\tau$ where spiral structure clearly appears.

We introduce numerical factor $C_{\rm g}$, which is defined as
\begin{equation}
	F_{\rm grav} =  C_{\rm g} \frac{\pi^3 G^2}{\Omega^2}r^2\Sigma^3.
							\label{Cg}
\end{equation} 
The estimation $F_{\rm grav}^{\rm est}$ for $i = 30$ degree corresponds to $C_{\rm g} = 0.38$.
We plot $C_{\rm g}$ at $r=0.7 R_{\rm Roche}$ in numerical results in Fig. \ref{Fgrav} (b).
When $\tau \gtrsim 0.1$, $C_{\rm g}$ is $\sim$ 1-2, independent of $\tau$ and $r_{\rm p}$, which indicates $F_{\rm grav}$ has the same functional form as $F_{\rm grav}^{\rm est}$.

\subsubsubsection{5.3.2.2 Translational Angular Momentum Transfer}
We show the relation between $F_{\rm trans}$ and $\Sigma$ at $r=0.7 r_{\rm Roche}$ in Fig. \ref{Ftrans} (a).
The dashed line is explained below.
In a similar way to equation (\ref{Cg}), we define a numerical factor $C_{\rm t}$ as
\begin{equation}
	F_{\rm trans} =  C_{\rm t} \frac{\pi^3 G^2}{\Omega^2}r^2\Sigma^3.
							\label{Ct}
\end{equation}
We show $C_{\rm t}$ in Fig \ref{Ftrans} (b).
Figs. \ref{Ftrans} show $F_{\rm trans}$ is always almost equal to $F_{\rm grav}$ at $0.7 R_{\rm Roche}$ where spiral structure develops.
In the outermost region near Roche limit, the results are more noisy in global simulation.
However, $C_{\rm g}$ and $C_{\rm t}$ seem to be larger by a factor of 2-3 near Roche limit in general.
$F_{\rm col}$ becomes less important with increase of $r$ (see Figs. \ref{AM}).
The local $N$-body simulation \citep{daisaka2001} also shows similar tendency on $r$.

In the outer region, $F_{\rm trans}$ is far greater than that corresponding to equation (\ref{FtransGT}) with $\sigma \sim r_{\rm p}\Omega$, which was derived under assumption of spatial uniformity by \citeauthor{goldreich1978}.
The development of spiral structure is responsible for the enhancement of $F_{\rm trans}$.
To see this more clearly, we separate $F_{\rm trans}$ into the component of bulk motion $F_{\rm bulk}$ and that of random motion $F_{\rm local}$, as explained in section 2.3.2.

In Figs. \ref{FbulkandFlocal}, we show $F_{\rm bulk}$ and $F_{\rm local}$ of RUNs in SET 3, averaged from $t = 4 T_{\rm k}$ to $6 T_{\rm k}$, as a function of $n_{\rm bulk}$.
The dashed lines are $\lambda_{\rm bulk}$, which is the scale of the region in which $n_{\rm bulk}$ particles exist.
Figs. \ref{FbulkandFlocal} (a) and (b) show $F_{\rm bulk}$ and $F_{\rm local}$ at $r = 0.7 R_{\rm Roche}$ for RUN 3-100K, RUN3-10K, respectively.
The predicted radial scale of spirals $\lambda_{\rm cr}$ (eq.[\ref{lambdaunstable}]) in the cases of Figs. \ref{FbulkandFlocal} (a) and (b) are about $0.08 R_{\rm Roche}$.
In Fig. \ref{FbulkandFlocal} (a), $F_{\rm bulk}$ and $F_{\rm local}$ are almost constant up to $n_{\rm bulk} \sim 100$, and $\lambda_{\rm bulk} \sim \lambda_{\rm cr}$.
This means that the particles move collectively as a group with scale $\sim \lambda_{\rm cr}$ in which about 100 particles exist.
In this scale, $F_{\rm local} \ll F_{\rm bulk}$, and translational angular momentum transfer is almost wholly due to this bulk motion.
This is the case also for RUN 2-10K (Fig. \ref{FbulkandFlocal}(b)).
In this case, particles move collectively with about 10 neighboring particles with the scale $\lambda_{\rm bulk}\sim\lambda_{\rm cr}$.
These results explain why $F_{\rm trans}$ is enhanced over equation (\ref{FtransGT}) and has the functional dependence predicted by equation (\ref{Ct}).

When the structure develops, $Q = \sigma \Omega / \pi G \Sigma \sim O(1)$.
Since $F_{\rm bulk} \gg F_{\rm local}$, $\sigma \sim v^{\rm bulk} \gg v^{\rm local}$.
Then, 
\begin{equation}
	v^{\rm bulk} \sim \frac{\pi G \Sigma}{\Omega} Q \sim \frac{\pi G \Sigma}{\Omega}.
                                                        \label{vbulkQ}
\end{equation}
Since particles move as groups, these groups may be treated as super particles with random velocity $v_{\rm bulk}$.
Life time of each structure is $\sim \Omega^{-1}$, so that collision frequency of the super particles is $\omega_{\rm c} \sim \Omega$.
Thus, translational viscosity and corresponding angular momentum flux may be 
\begin{equation}
	\nu_{\rm trans}^{\rm est} \sim \frac{(v_r^{\rm bulk})^2}{\Omega} \sim \frac{\pi^2 G^2 \Sigma^2}{\Omega^3},
							\label{nutransest}
\end{equation}
\begin{equation}
	F_{\rm trans}^{\rm est} \sim 3 \frac{\pi^3 G^2}{\Omega^2} r^2 \Sigma^3.
							\label{Ftransest}
\end{equation}
Note that $F_{\rm trans}^{\rm est} \sim F_{\rm grav}^{\rm est}$ and $\nu_{\rm trans}^{\rm est} \sim \nu_{\rm grav}^{\rm est}$.
We also plot $F_{\rm trans}^{\rm est}$ in Fig. \ref{Ftrans} (a) with a dashed line.
Considering a disk with turbulence induced by self-gravitational instability, \nocite{ward1978}Ward \& Cameron (1978) and \nocite{lin1987}Lin \& Pringle (1987) derived a similar angular momentum transfer rate.

In parameters of Saturn's ring, \nocite{daisaka2001}Daisaka et al. (2001) studied more detailed relation between angular momentum fluxes and $\tau$, $\Sigma$, and $r$ by local simulations.

\subsubsection{Angular momentum transfer in higher restitution coefficient case}

We have studied the cases with $\varepsilon_{\rm n} = 0.1$.
This choice may lead to rather high cooling rate of random velocity, keeping $Q$ value low.
Equilibrium random velocity is determined by heating due to transfer of shear motion to random motion through collisions and damping due to inelastic collisions (\citeauthor{goldreich1978}; \citealt{ohtsuki1993}).
However, for $\varepsilon_{\rm n} > 0.67$, the damping is no longer strong enough to attain equilibrium state and random velocity keeps growing (\citeauthor{goldreich1978}).
Then, $Q$ value should be so high that the spiral structure does not appear.
To examine the effect of higher $\varepsilon_{\rm n}$, we performed additional simulations (SET6) with restitution coefficient 0.2, 0.4, 0.6 and 0.8.
We show the snapshots of some cases in SET6 in Figs. \ref{snaprestitutions}. 
Initial surface density distribution is the same as that in SET3.
Panel (a) is the case with $\varepsilon_{\rm n} = 0.4$ at $t=8 T_{\rm K}$.
Panel (b) is the case with $\varepsilon_{\rm n} = 0.6$.
For $\varepsilon_{\rm n} \lesssim 0.4$, spiral structure seems similar to the runs with $\varepsilon_{\rm n} = 0.1$.
However, when $\varepsilon_{\rm n} = 0.6$, the spiral is less clear.
In the case with $\varepsilon_{\rm n} = 0.8$, any structure does not appear.
We show angular momentum transfer rates in Figs. \ref{fAMrestitution}. 
These rates are averaged over $t=6 T_{\rm K}$ to $t=8 T_{\rm K}$.
For $\varepsilon_{\rm n} = 0.6$, gravitational transfer is less dominant.
This reflects the less clear spiral structure (Fig. \ref{fAMrestitution} (b)).

We show $F_{\rm grav}$ as a function of $\Sigma$ for different $\varepsilon_{\rm n}$ in Figs. \ref{Cgrestitution} (a).
Only data with optical depth larger than 0.3 is chosen.
Dashed lines are the best fitted lines with assumption that $F_{\rm grav} \propto \Sigma^3$.
While $F_{\rm grav}$ decreases with increase of $\varepsilon_{\rm n}$, the relation $F_{\rm grav} \propto \Sigma^3$ remains.
$F_{\rm trans}$ and $F_{\rm col}$ are plotted in panels (b) and (c).
Although they decrease with increase of $\varepsilon_{\rm n}$, the decrease is smaller than that in $F_{\rm grav}$.
($F_{\rm trans}$ does not show clear dependence as $\Sigma^3$ in large $\varepsilon_{\rm n}$ case.
We omit the fitted line in the case with $\varepsilon_{\rm n} = 0.8$ because of large dispersion.)

Panel (d) shows the corresponding $C_{\rm g}$, $C_{\rm t}$ and $C_{\rm c}$ as a function of $\varepsilon_{\rm n}$.
Error bars represent standard deviations.
In general, as $\varepsilon_{\rm n}$ increases, angular momentum transfer, in particular $F_{\rm grav}$, decreases.
However, the results do not change so much ($C_{\rm g}, C_{\rm c}$ and $C_{\rm t}$ change only by factor 2) except for highly elastic case $\varepsilon_{\rm n} \gtrsim 0.6$, when the spiral structure is not clear.

\subsubsection{Summary of Angular Momentum Transfer}

In an optically thick particle disk, angular momentum is transferred in different ways in inner and outer regions.
The boundary between inner and outer regions is about 0.6-0.7 $R_{\rm Roche}$.
In the inner region, clear spiral structure does not develop, so that the analytical formulae assuming spatial uniformity work well (e.g. \citeauthor{goldreich1978,araki1986}).
In this region, collisional angular momentum transfer dominates, which is well fitted by equation (\ref{nucolnumerical}).

In the outer region, $F_{\rm grav}$ and $F_{\rm trans}$ are enhanced by the spiral structure and dominate $F_{\rm col}$.
They are both proportional to $\Sigma^3$ and independent of $r_{\rm p}$.
In outer region, $F_{\rm grav} \simeq F_{\rm trans} > F_{\rm col}$.
Defining $C_i (i={\rm g,t,c})$ as $F_{\rm grav}$, $F_{\rm trans}$ and $F_{\rm col} = C_i (\pi^2 G^2/\Omega^2)r^2 \Sigma^3 \; (i = {\rm g,t,c})$, our numerical results show $C_{\rm total} = C_{\rm g} + C_{\rm t} + C_{\rm c} \sim C_{\rm g} + C_{\rm t} \sim$ 4-8.
In local simulations, $C_{\rm total} \sim 2$ in the region corresponding to $r \sim 0.7 R_{\rm Roche}$ and $C_{\rm total} \sim 6$ at $r \sim 0.9 R_{\rm Roche}$ \citep{daisaka2001}.
The numerical factor slightly increases with $r$ in local simulations \citep{daisaka2001}.
Similar tendency is found in our global simulations.
As the restitution coefficient increases, numerical factors decrease and relative importance of $F_{\rm grav}$ diminishes.
However, the transfer rate changes only by a factor 2 unless $\varepsilon_{\rm n}$ is highly elastic ($\varepsilon_{\rm n} \gtrsim 0.6$).

We show time evolution of surface density distribution in log scale in Figs. \ref{surdenlog}, for (a) RUN 3-100K and (b) RUN 5-30K.
Initial surface density distribution is proportional to $\Sigma \propto r^{-3}$ in RUN 3-100K and flat in RUN 5-30K.
Dashed lines in the figures are proportional to $\Sigma^{\rm -3/2}$.
Surface density distribution after initial relaxation but before formation of lunar seeds in our simulations is consistent with $\Sigma \propto r^{-3/2}$ in outer region.
The relation $\Sigma \propto r^{-3/2}$ is the steady accretion solution ($d\dot{M}_{\rm disk}/dr = 0$) to equation (\ref{massflowhikakuwakusei}) with constant $C_{\rm total}$ \citep{lin1987}.

\section{CONCLUSIONS AND DISCUSSION}
	
We have investigated angular momentum transfer and associated mass transfer in a particle disk where physical collisions as well as self-gravity are important.
First we presented formulation of angular momentum transfer in the disk, starting from Boltzmann equation.
Next, we performed global N-body simulation with $N = 10^3$-$10^5$ to directly calculate angular momentum transfer fluxes, based on the above formulation.
We simulated disks that correspond to a protolunar disk generated by a giant impact on the proto Earth.
The disk has total mass $\sim 0.02$ - $0.06 M_{\rm c}$ where $M_{\rm c}$ is the central body mass and most mass is initially within the Roche limit of the central body.
In such a dense disk, spiral structure is formed by self-gravity and energy dissipation due to inelastic collisions, except for innermost region where tidal force of the central body is too strong.

In the region $0.6 R_{\rm Roche} \lesssim r \lesssim 1 R_{\rm Roche}$, angular momentum transfer is regulated by gravitational torque exerted by spiral structure and collective motion of particles in the spirals.
\citeauthor{ida1997} and \citeauthor{kokubo2000} showed that formation of the Moon is regulated by angular momentum transfer in this region.
With increasing $\tau$, spiral structure becomes clearer.
When $\tau \gtrsim 0.2$, angular transfer rate is
\begin{equation}
	F_{\rm grav} + F_{\rm trans} + F_{\rm col} \sim F_{\rm grav} + F_{\rm trans} = (C_{\rm g} + C_{\rm t}) \frac{\pi^3 G^2}{\Omega^2}r^2 \Sigma^3,
							\label{Fconclusion}
\end{equation}
where $C_{\rm g} + C_{\rm t}$ is about 4-8 in the parameters for a protolunar disk. 
Then, surface density distribution approaches to distribution in steady accretion that is proportional to $r^{-3/2}$.

The average optical depth is
\begin{equation}
	\tau=\frac{N \pi r_p^2}{\pi(R_{\rm Roche}^2 - R_\oplus^2)} = \left( \frac{3 M_{\rm disk}}{4\pi\rho} \right)^{2/3} \frac{1}{(R_{\rm Roche}^2 - R_\oplus^2)} N^{1/3}.
\end{equation}
Since $C_{\rm g} + C_{\rm t}$ saturates for $\tau \gtrsim 0.2$, particle number $N$ in global simulation must be
\begin{equation}
	N \gtrsim 3500 \left( \frac{M_{\rm disk}}{3 M_L} \right)^{-2}.
							\label{Ncondition}
\end{equation}
The relation between the initial disk mass and the mass of the Moon was derived by \citeauthor{ida1997}.
When initial mass is distributed within Roche limit, required initial disk mass is about 3 times the present lunar mass.
Thus, evolution of the protolunar disk can be followed with rather small number of particles; $N \gtrsim 3000$ is enough and $N \gtrsim 1000$ ($\tau \gtrsim 0.15$) may be okay.

In the innermost region, spiral structure does not develop and angular momentum transfer is dominated by collisions between particles, and the corresponding viscosity has positive dependency on particle size.
N-body simulations with limited number overestimate the diffusion process in this region.
However, this does not change the lunar formation, since lunar seed is formed by the angular momentum transfer in outer region.
Once a large proto-moon is formed, the remaining disk would interact with the lunar seed, and the disk materials would eventually be scattered to fall to the Earth (\citeauthor{ida1997,kokubo2000}).
Thus, we conclude that N-body simulation of evolution of a protolunar disk is not affected by limited number of particles, as long as $N \gtrsim$ 1000-3000.

The time scale of viscous evolution is given as $\sim \Delta r^2/\nu$.
Since total viscosity is 
\begin{equation}
	\nu_{\rm total} = C_{\rm total}\frac{\pi^2 G^2}{3 \Omega^3} \Sigma^2,
\end{equation}
the time scale of disk evolution is 
\begin{equation}
	t_{\rm ev} \simeq \frac{500}{C_{\rm total}} \left( \frac{\Sigma}{0.01 M_\oplus R_{\rm Roche}^{-2}} \right)^{-2} \left( \frac{\Delta r}{R_{\rm Roche}} \right)^2 \left( \frac{r}{R_{\rm Roche}} \right)^{-9/2} T_{\rm K}.
\end{equation}
Since $C_{\rm total} \simeq C_{\rm g} + C_{\rm t}$ is about 4-8, $t_{\rm ev} \sim 100 T_{\rm K}$, which is consistent with lunar formation time obtained by \citeauthor{ida1997} and \citeauthor{kokubo2000}.
The development of spiral structure is essential for the rapid evolution of a protolunar disk.

We comment on the physical processes we neglected and validity of assumptions in our simulations.
One process we neglected is fragmentation.
Since typical random velocity is large, catastrophic disruption would occur in particle collisions.
Thus, realistic particle size would be much smaller.
However as long as random energy damping is sufficient, spiral spiral structure develops as well.
Thus fragmentation would not affect the evolution of the disk, since the angular momentum transfer in the outer region has no dependency on particle size, being regulated by the movement of particles as a group.

We simply assumed that all particles have the same size, so that filling factor is 0.7 at most.
If size distribution is included, the effective material density increases.
It may expand the region where clear spiral develops to more inner region.
Also, random velocity of smaller particles would be larger if size distribution is included. 
However, size distribution does not prevent the spiral structure developing in the simulations in (\citeauthor{ida1997} and \citeauthor{kokubo2000}), so that this would not affect the physics in principle.

We also assumed that the central body is spherical.
Objects interact with tidal bulges raised on the Earth, so that materials exterior to the synchronous orbit migrate outward, and interior materials migrate inward.
The time scale of tidal evolution of a body with one lunar mass is of order of $10^3$ yr \citep{canup1996}.
This is much longer than the time scale of disk evolution and the tidal effect is not essential for disk evolution.
The formed moon would migrate outward sweeping up remnants of the disk \citep{canup1999}.
The Earth itself may be deformed considerably without tidal bulges.
In SPH simulations of giant impact, the core of impactor penetrates through the mantle material and accumulates on proto-Earth, and forms a rotating quadrupole (e.g. \citealt{cameron1997}).
Though the quadrupole subsides substantially in a day or two, some fraction of quadrupole component may remain.
This would not effect the angular momentum transfer due to local instability, which was discussed in this paper, since the time scale of angular momentum transfer by local instability is very short.
However, this may have considerable effect in a longer time scale.

We also assumed that a protolunar disk is an entirely particulate disk.
In recent SPH simulations \citep{cameron1997} suggests that vapor/liquid and solid phase coexist in a protolunar disk after the giant impact, with average temperature above 4000K.
Assuming that an initial disk with 3 lunar mass is within Roche limit, heat dissipation required for the disk evolution is about $1.5 \times 10^{37} $erg, which is about a half of latent heat of vaporization of silicate of the disk mass.
If an entire disk is evaporated, Toomre's Q value is $\sim 3 (T/1000{\rm K})^{1/2}(r/R_{\rm Roche})^{-3/2}[\Sigma/(0.01M_\oplus/R_{\rm Roche}^2)]^{-1}$, with mean molecular weight 30, and the disk would be stable against gravitational instability.
During a short time scale of dynamical disk evolution, it is difficult for radiation to cool down the disk \citep{thompson1988}.
If gravitational instability is aborted by the vaporization, the disk evolution would be regulated by the cooling time of the disk \citep{thompson1988}.	
The disk evolution and Moon accretion would depend on how the disk with vapor/liquid-solid mixture evolves, as discussed in \citeauthor{kokubo2000}.
If the disk remains within Roche limit until the disk is sufficiently cooled down to develop gravitational instability, a single moon would be formed just beyond Roche limit.
The disk would be condensed from outer region.
Spirals would develop in the condensed region, which results in rapid diffusion of the materials there.
A single moon accretes the materials diffused out, staying at the location just beyond Roche limit.
The eventual mass of the moon would be the same as the results of N-body simulations neglecting vaporization, because final moon mass is determined by redistribution of disk angular momentum to the moon and materials that fall to the Earth (\citeauthor{ida1997,kokubo2000}), although time scale may be regulated by cooling.

The heat generation problem occurs in a very heavy disk such as a protolunar disk.
The result that angular momentum transfer in a particulate disk within Roche limit is regulated by gravitational instability is applicable to other disk systems, such as planetary rings or small satellite formation.
In planetary rings, the only difference is that the wave length of structure $\lambda_{\rm cr}$ is much smaller since $M_{\rm disk}/M_{\rm c} \ll 1$ (see eq. [\ref{lambdaunstable2}]).
\citet{daisaka2001} performed the local N-body simulations with the parameters of Saturn's B-ring, and calculated the angular momentum transfer rate in a similar way.
They found that the angular momentum transfer in Saturn's B-ring is also regulated by wake like structure and equation (\ref{Fconclusion}) also holds.

\acknowledgments

We are grateful for helpful comments by Hidekazu Tanaka on formalizations.
Discussion with Hiroshi Daisaka was valuable.
We also thank his technical comments on numerical calculations.
The numerical calculations were carried out on a special purpose computer \mbox{GRAPE-4}.
We thank Junichro Makino and Eiichro Kokubo for technical advice on \mbox{GRAPE-4}.
We also thank the anonymous referee for useful comments.
This work was supported by \mbox{Grant-in-Aid} for Scientific Research (c) 12640405.

\appendix
\section{ANGULAR MOMENTUM FLUX DUE TO GRAVITATIONAL TORQUE}

A formula for angular momentum flux due to gravitational torque with a spiral pattern is given by \citet{lynden1972}.
From the definition (eq. [\ref{DefFgrav}]), 
\begin{equation}
	F_{\rm grav} = \int_{r_{\rm min}}^r dr' \int_0^{2\pi} r'd\theta 
	\int_{-\infty}^{\infty} dz \,
		\rho \pderiv{\Phi'}{\theta},
						\label{potentialFgrav}
\end{equation}
where $\Phi'$ is potential due to disk material.
Combining equation (\ref{potentialFgrav}) with Poisson equation, it is expressed as 
\begin{equation}	
	F_{\rm grav}(r) = \frac{1}{4 \pi G} 
		\int_0^{2\pi} rd\theta \int_{-\infty}^{\infty} dz
	\pderiv{\Phi'}{r} \pderiv{\Phi'}{\theta}.
						\label{Fgravcontinuous}
\end{equation}
We represent surface density distribution as
\begin{equation}
	\Sigma(r,\theta,t) = \Sigma_0(r,t) + \Sigma_1(r,\theta,t) = \Sigma_0(r,t) + H(r,t)e^{m\theta + f_{\rm s}(r,t)},
						\label{Sigma}
\end{equation}
where $\Sigma_0$ is azimuthally averaged density distribution and $f_s(r,t)$ is shape function for the $n$th arm satisfying
\begin{equation}
	n \Phi' + f_s(r,t) = {\rm constant \; (mod \; 2\pi)}.
						\label{shapefunction}
\end{equation}
Using tightly winding approximation (e.g. \citealt{binney1987}), the potential due to $\Sigma_1$ is given by 
\begin{equation}
	\Phi'_1(r,\theta,z,t) = - \frac{2 \pi G}{|k|} H(r,t)
		{\rm Re} \left\{ e^{i \{m \theta + f_s(r,t) \} - |kz| } \right\},
						\label{potential}
\end{equation}
where $k$ is wave number.
Note that $r$ component of $k$ is $k_r = \partial f_s / \partial r = k \cos i$.
Substituting equation (\ref{potential}) into equation (\ref{Fgravcontinuous}), and using $m = k_r r \tan i$, where $i$ is pitch angle, we obtain 
\begin{equation}
	F_{\rm grav} = \frac{\pi^2 G r}{k^2} \frac{k_{\rm r}}{k} m H^2 = \frac{\pi}{2}Gr^2\lambda_r \sin i \cos^2 i,
						\label{Fgravest0}
\end{equation}
where $\lambda_r = 2\pi/k_r$.

\newpage

\figcaption{
Snapshots of RUN 2-100K.
(a), (b), (c), (d), (e) and (f) are snapshots at $t=0,2,6,10,20,30 T_{\rm K}$, respectively.
The inner circle is the Earth with radius $0.343 R_{\rm Roche}$.
The large outer circle shows the Roche limit.
\label{snap}}

\figcaption{
Snapshots of RUN 2-100K on $r$-$z$ plane, corresponding to Figures \ref{snap}.
Three circles show Earth's radius, 0.5 $R_{\rm Roche}$, and $1 R_{\rm Roche}$.
\label{snapedge}}

\figcaption{
Snapshots of (a) RUN 1-100K, (b) 2-100K, and (c) 3-100K at $t=6 T_{\rm K}$ and (d) 4-100K at $t=4 T_{\rm K}$.
\label{snapcompare}}

\figcaption{
Time evolution of surface density in (a) RUN 2-100K and (b) RUN 5-30K.
Surface density profile at (c) $t=6 T_{\rm K}$ and (d) $t=18 T_{\rm K}$ in  SET 2. 
The Earth's radius is at 0.343 $R_{\rm Roche}$.
\label{surden1}}

\figcaption{
Magnified snapshots for (a) RUN 2-100K and (b) RUN 2-10K at $t=10 T_{\rm K}$, (c) RUN 3-100K and (d) RUN 3-10K at $t=6 T_{\rm K}$.
The three circles are the radius of Earth, $0.5 R_{\rm Roche}$ and $1R_{\rm Roche}$.
\label{closeup}}

\figcaption{
Contours of autocorrelation ${\rm Corr}(\Delta r, \Delta\theta)$ at $r = 0.7 R_{\rm Roche}$, with different surface density, (a) $\Sigma = 0.005$-$0.006$, (b) $\Sigma = 0.008$-$0.010$, and (c) $\Sigma = 0.012$-$0.016$.
Panel (d), (e), and (f) are the results with (d) $ r=0.5 R_{\rm Roche}$ and $\Sigma = 0.012$-$0.016$, (e) $r=0.6 R_{\rm Roche}$ and $\Sigma = 0.012$-$0.016$, and (c) $r=0.9 R_{\rm Roche}$ and $\Sigma = 0.004$-$0.005$.
\label{figcorrelation}}

\figcaption{
Angular momentum fluxes $F_{\rm col}$, $F_{\rm grav}$ and $F_{\rm trans}$ as functions of $r$, averaged over $t=4$-$6 T_{\rm K}$.
(a) RUN 2-100K,
(b) RUN 2-10K,
(c) RUN 3-100K, and (d) RUN3-10K.
\label{AM}}

\figcaption{
$Q$ value as a function of $r$, in (a) SET 2 and (b) SET 3 at $t=6 T_{\rm K}$.
\label{Qvalue}}

\figcaption{
Collisional viscosity $\nu_{\rm col}/\Omega r_{\rm p}^2$ is given as a function of $\tau$, in the region where spiral structure does not develop.
Circles are the results of $N$-body simulations at $r=0.5 R_{\rm Roche}$ for different times and RUNs.
A dashed line is a fitted value given by equation (\ref{nucolnumerical}).
The solid line is analytical estimate by GT78 and AT86.
\label{nucoltau}}

\figcaption{
(a) $F_{\rm grav}$ as a function of surface density $\Sigma$ at $r=0.7 R_{\rm Roche}$.
The dashed line shows the estimated $F_{\rm garv}$ which is given as equation (\ref{Fgravorderest}) with $i=30$ degree.
(b) $C_{\rm g}$ as a function of optical depth $\tau$ at $r=0.7 R_{\rm Roche}$.
\label{Fgrav}}

\figcaption{
(a) $F_{\rm trans}$ as a function of surface density $\Sigma$ at $r=0.7 R_{\rm Roche}$.
The dashed line is $F_{\rm trans}^{\rm est}$ (eq. [\ref{Ftransest}]).
(b) $C_{\rm t}$ as a function of optical depth $\tau$ at $r=0.7 R_{\rm Roche}$.
\label{Ftrans}}

\figcaption{
$F_{\rm bulk}$ and $F_{\rm local}$ as a function of $n_{\rm bulk}$ at $r=0.7 R_{\rm Roche}$ in (a) RUN 3-100K and (b) RUN 3-10K.
Filled circles are $F_{\rm bulk}$ and circles are $F_{\rm local}$.
Dashed lines are $\lambda_{\rm bulk}$ corresponding to $n_{\rm bulk}$.
\label{FbulkandFlocal}}

\figcaption{
Snapshots for SET 6, with (a) $\varepsilon_{\rm n} = 0.4$ and (b) $\varepsilon_{\rm n} = 0.6$.
Initial surface density is similar to SET 3.
\label{snaprestitutions}}

\figcaption{
Angular momentum fluxes $F_{\rm col}$, $F_{\rm grav}$ and $F_{\rm trans}$ as functions of $r$, with different $\varepsilon_{\rm n}$.
(a) $\varepsilon = 0.4$ and (b) $\varepsilon = 0.6$. 
\label{fAMrestitution}}

\figcaption{
Angular momentum fluxes $F_{\rm col}$, $F_{\rm grav}$ and $F_{\rm trans}$ as function of $\Sigma$, with different $\varepsilon_{\rm n}$.
Dashed lines are best fitted line assumed that fluxes are proportional to $\Sigma^3$.
(a) $F_{\rm grav}$, (b) $F_{\rm trans}$, and (c) $F_{\rm col}$.
Dashed line in panel (b) is fitted for $\varepsilon_{\rm n} = 0.1$.
(d) is relation between $\varepsilon_{\rm n}$ and corresponding $C_{\rm g}$ and $C_{\rm c}$.
\label{Cgrestitution}}

\figcaption{
Time evolution of surface density $\Sigma$ in (a) RUN 3-100K and (b) RUN 5-30K.
Dashed lines are proportional to $r^{-3/2}$.
\label{surdenlog}}

\clearpage

\begin{deluxetable}{lcccccc}
\tablewidth{0pt}
\tablecaption{Parameters of initial disks for SET 1-6\label{initialdisk}}
\tablehead{
\colhead{} & \colhead{Disk Mass} & \colhead{} & \colhead{$a_{\rm min}$} & $a_{\rm max}$ & \colhead{} & \colhead{} \\ \colhead{SET} & \colhead{($M_{\rm L}$)\tablenotemark{a}} & \colhead{$\alpha$} & \colhead{$(R_{\rm Roche})$} & \colhead{$(R_{\rm Roche})$} & \colhead{$\langle e^2 \rangle^{1/2}$} & \colhead{$\langle i^2 \rangle^{1/2}$} }
\startdata
SET1 & 1.67 & -3 & 0.4 & 1.2 & 0.05 & 0.05 \\
SET2 & 2.47 & -3 & 0.4 & 1.1 & 0.15 & 0.3 \\
SET3 & 3.20 & -3 & 0.4 & 1.1 & 0.05 & 0.05 \\
SET4 & 4.94 & -3 & 0.4 & 1.1 & 0.05 & 0.05 \\
SET5 & 3.20 & 0 & 0.4 & 1.1 & 0.05 & 0.05 \\
SET6 & ...\tablenotemark{b} & -3 & 0.4 & 1.1 & 0.05 & 0.05\\ 
\enddata
\tablenotetext{a}{$M_{\rm L} = 0.0125M_\oplus = 7.35\times10^{25}$g}
\tablenotetext{b}{For each $\varepsilon_{\rm n} = 0.2, 0.4, 0.6$ and $0.8$, we performed four RUNs with disk mass 1.67, 2.47, 3.20 and 4.94.}
\end{deluxetable}

\clearpage

\begin{figure}
\plotone{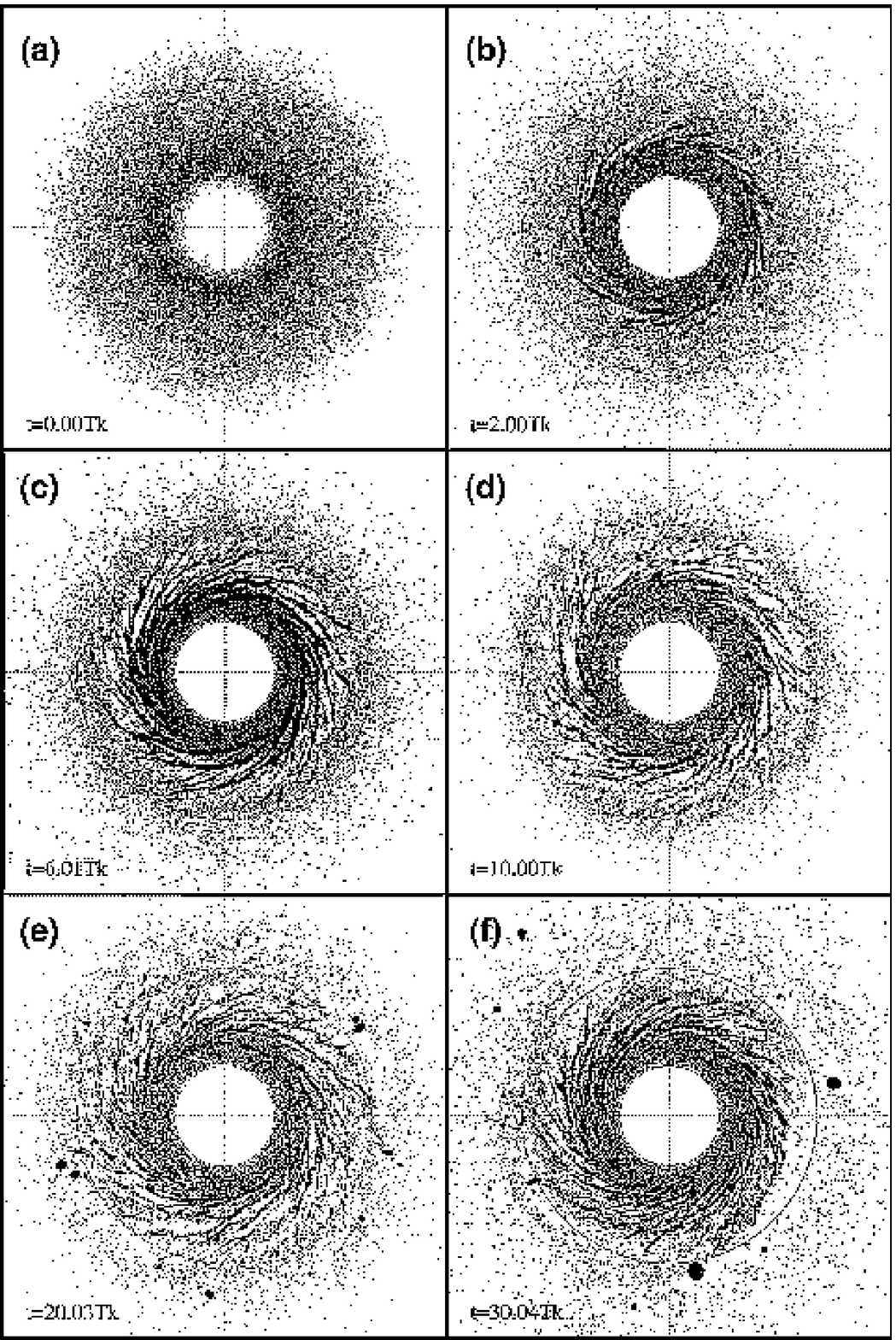}

Figs. \ref{snap}
\end{figure}

\begin{figure}
\plotone{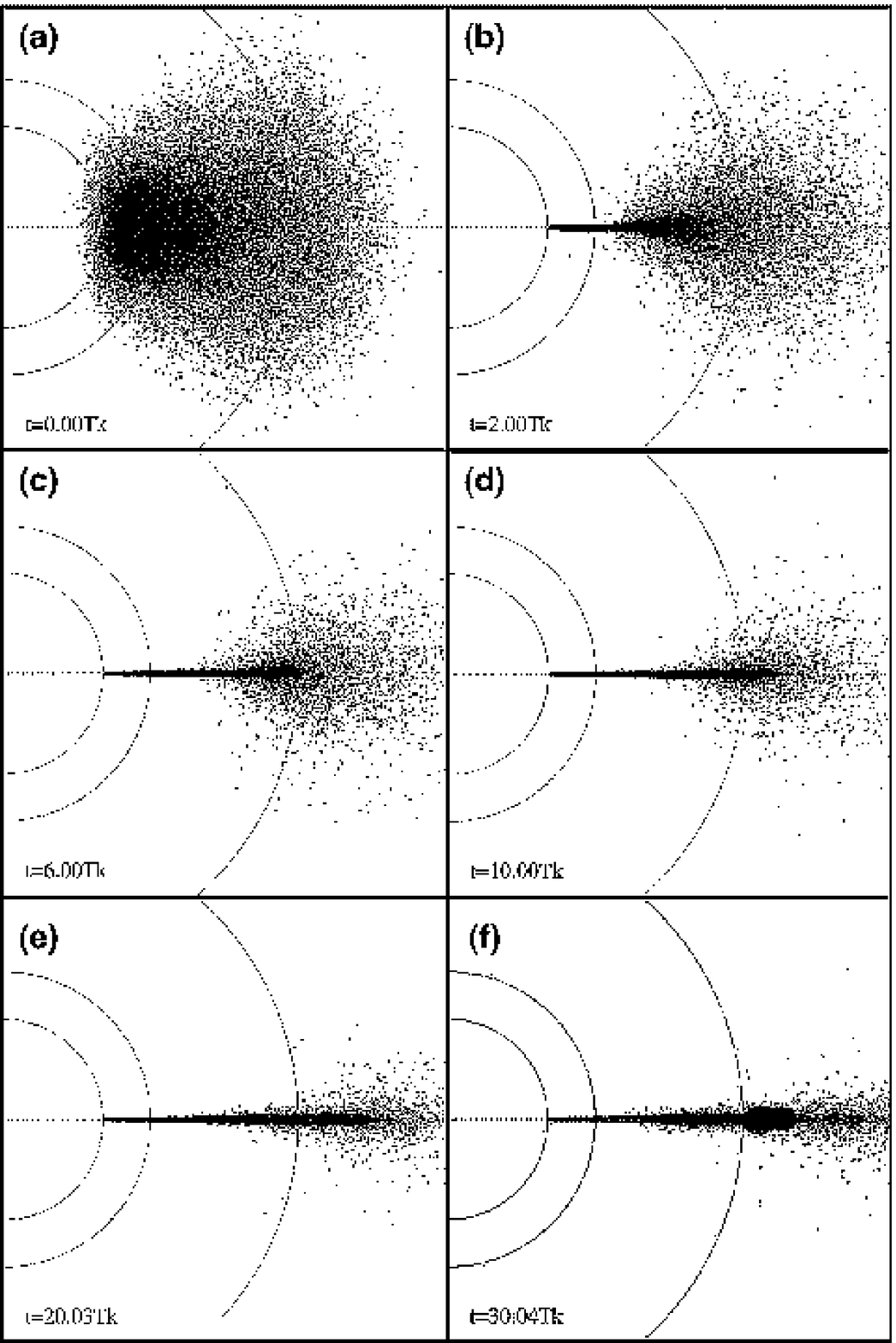}

Figs. \ref{snapedge}
\end{figure}

\begin{figure}
\plotone{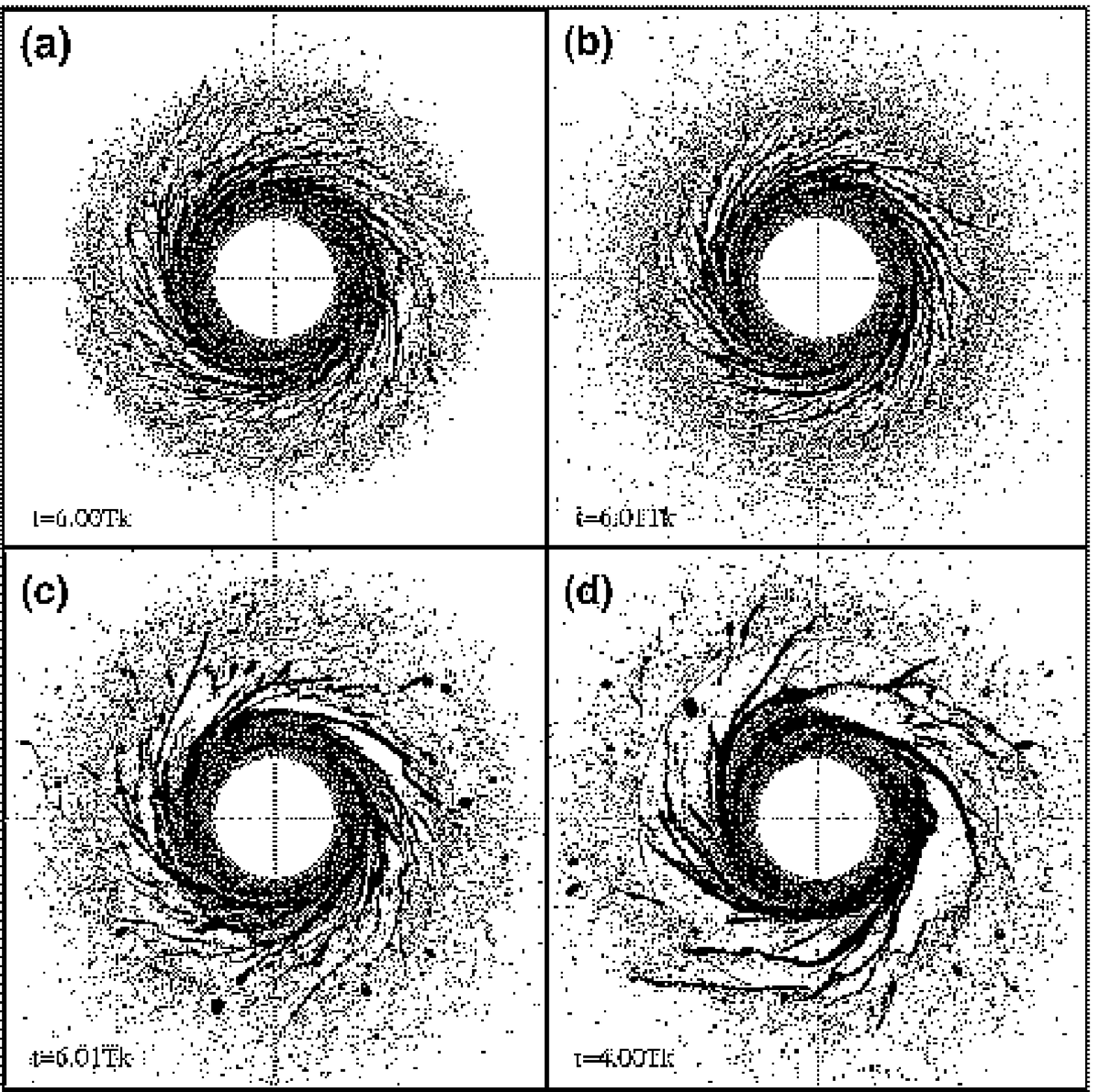}

Figs. \ref{snapcompare}
\end{figure}

\begin{figure}
\plotone{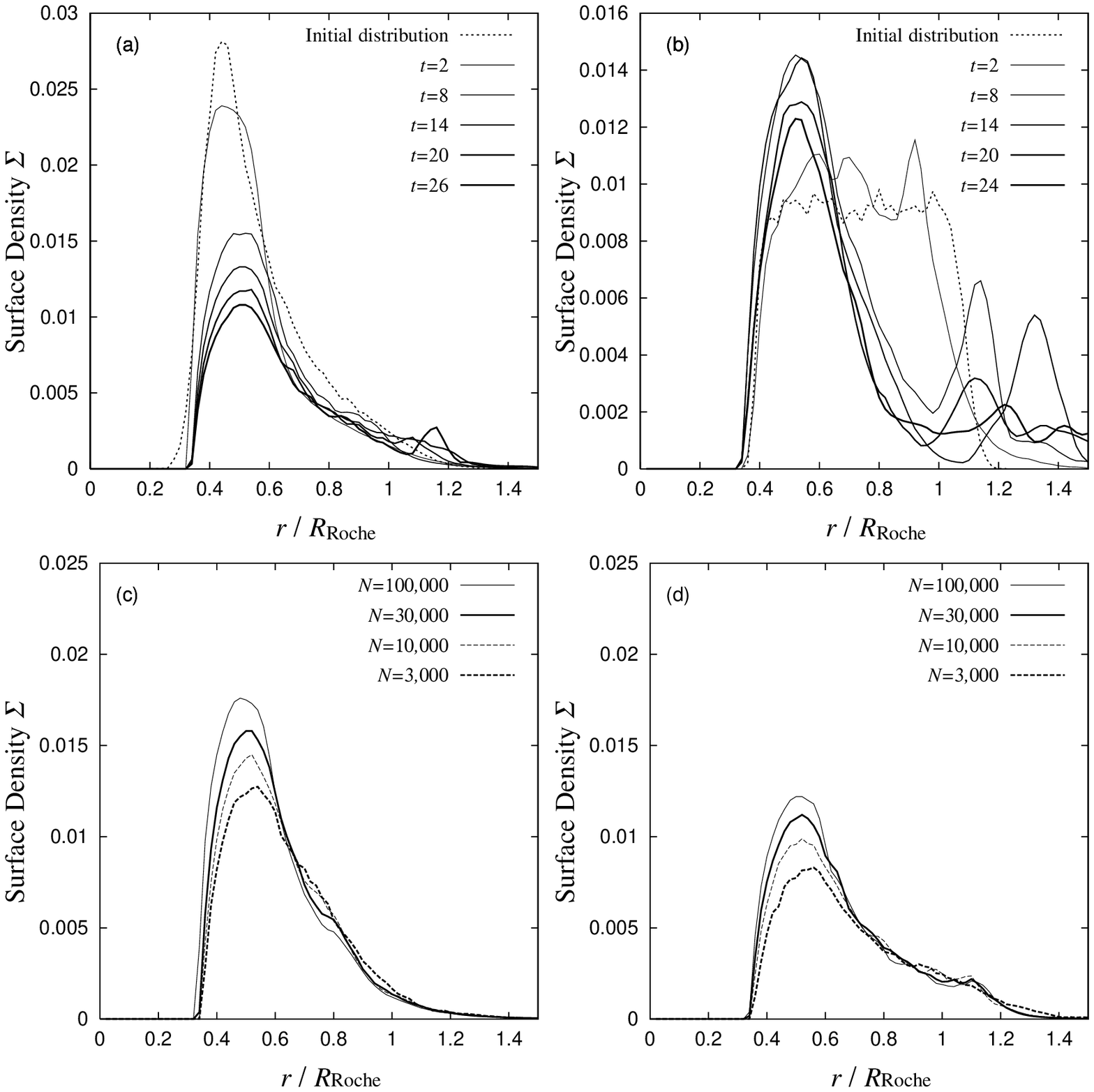}

Figs.\ref{surden1}
\end{figure}

\clearpage

\begin{figure}
\plotone{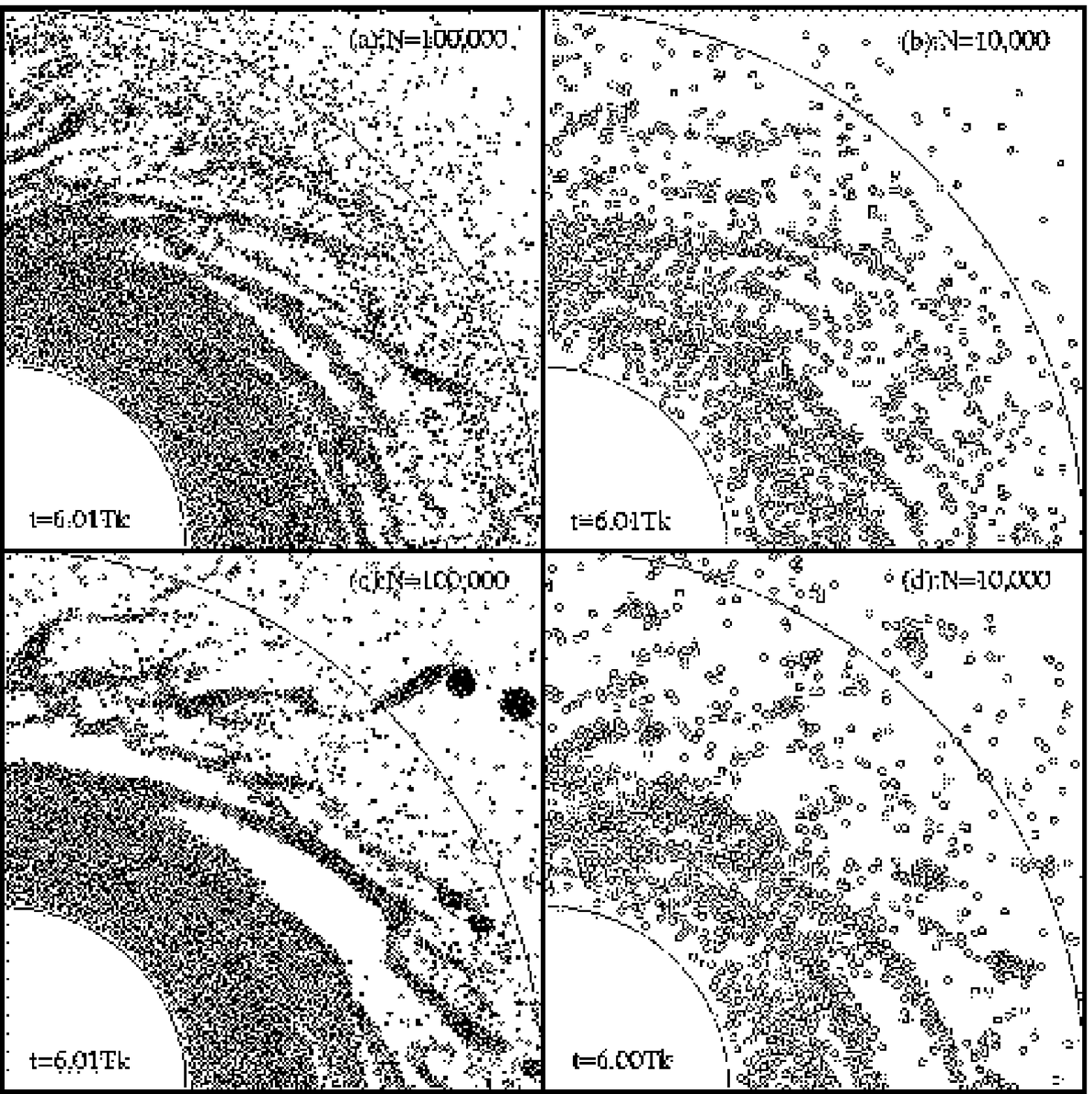}

Figs. \ref{closeup}
\end{figure}

\begin{figure}
\plotone{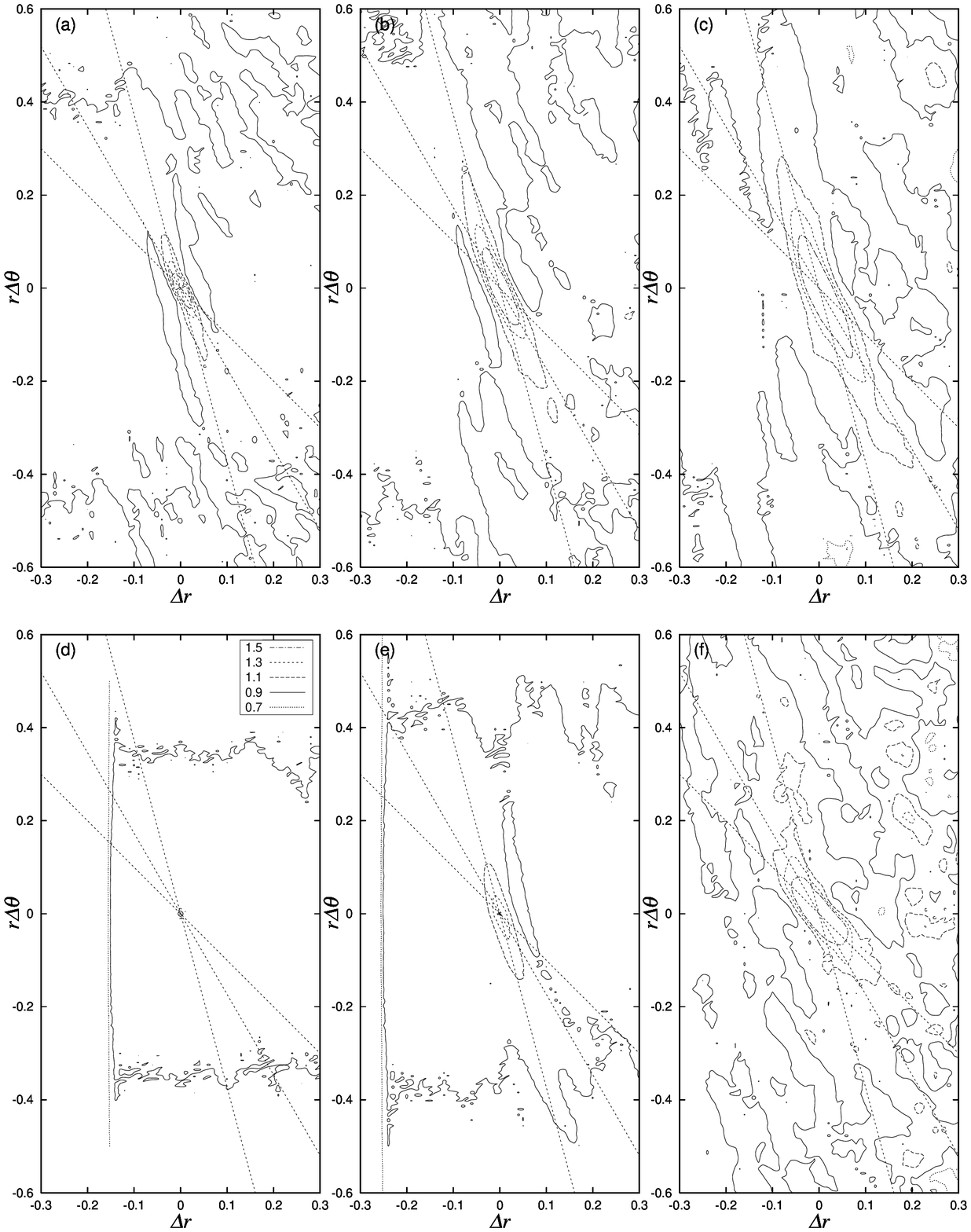}

Figs. \ref{figcorrelation}
\end{figure}

\begin{figure}
\plotone{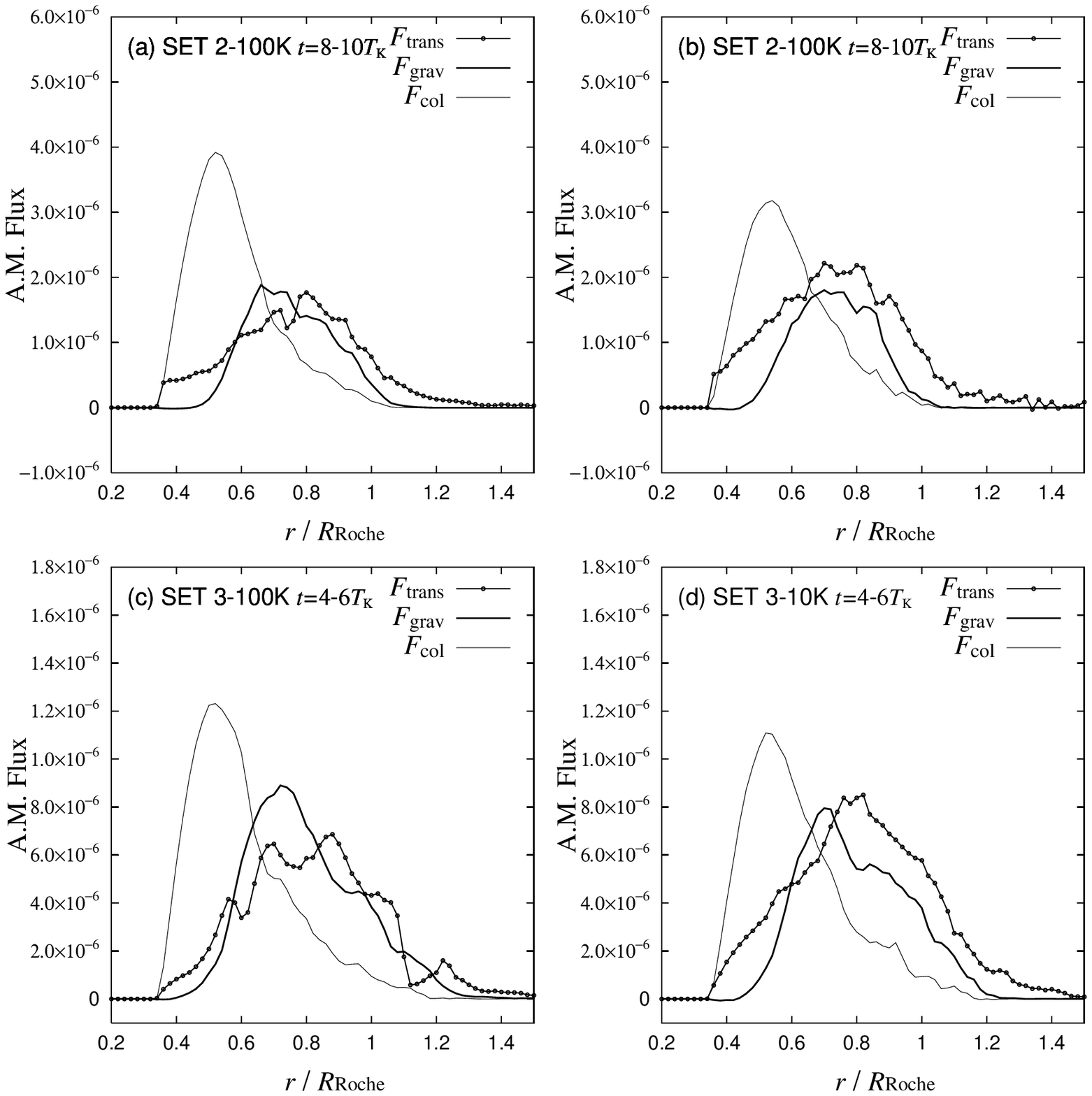}

Figs.\ref{AM}
\end{figure}

\begin{figure}
\plotone{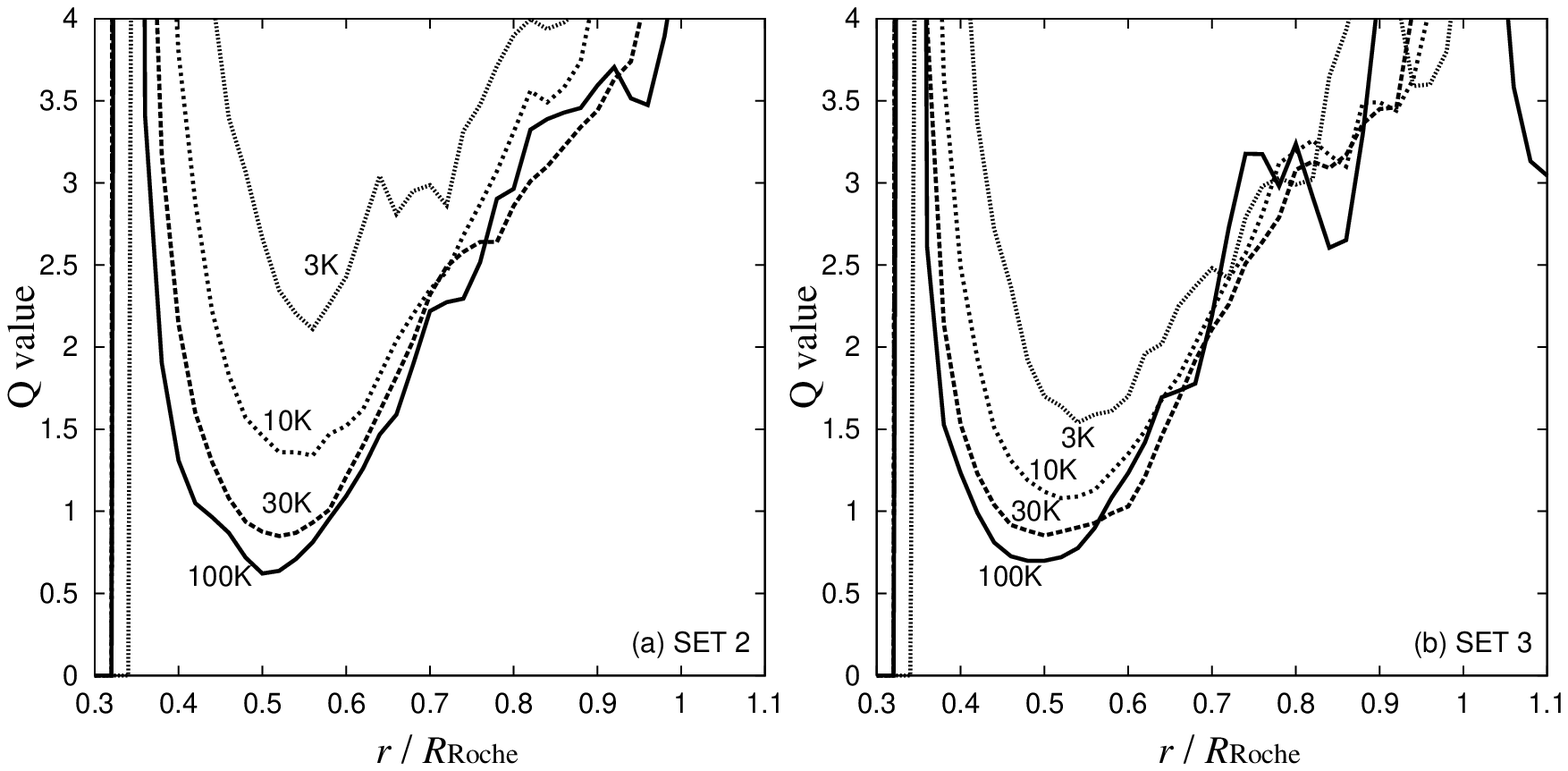}

Figs. \ref{Qvalue}
\end{figure}

\begin{figure}
\plotone{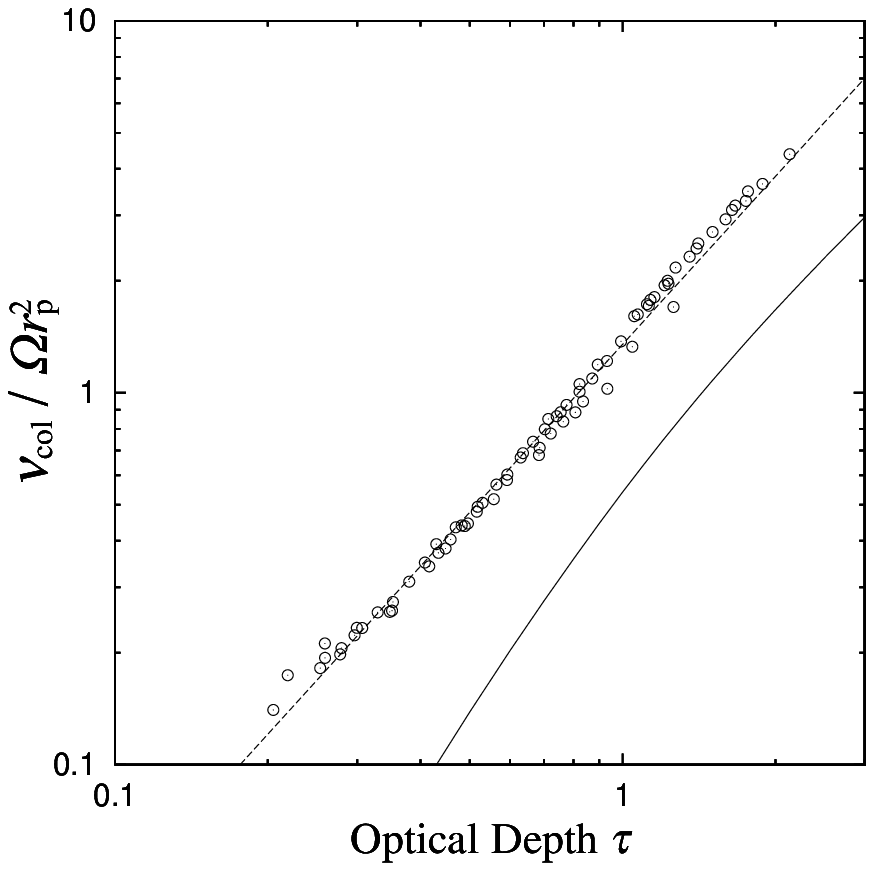}

Fig.\ref{nucoltau}
\end{figure}

\begin{figure}
\plotone{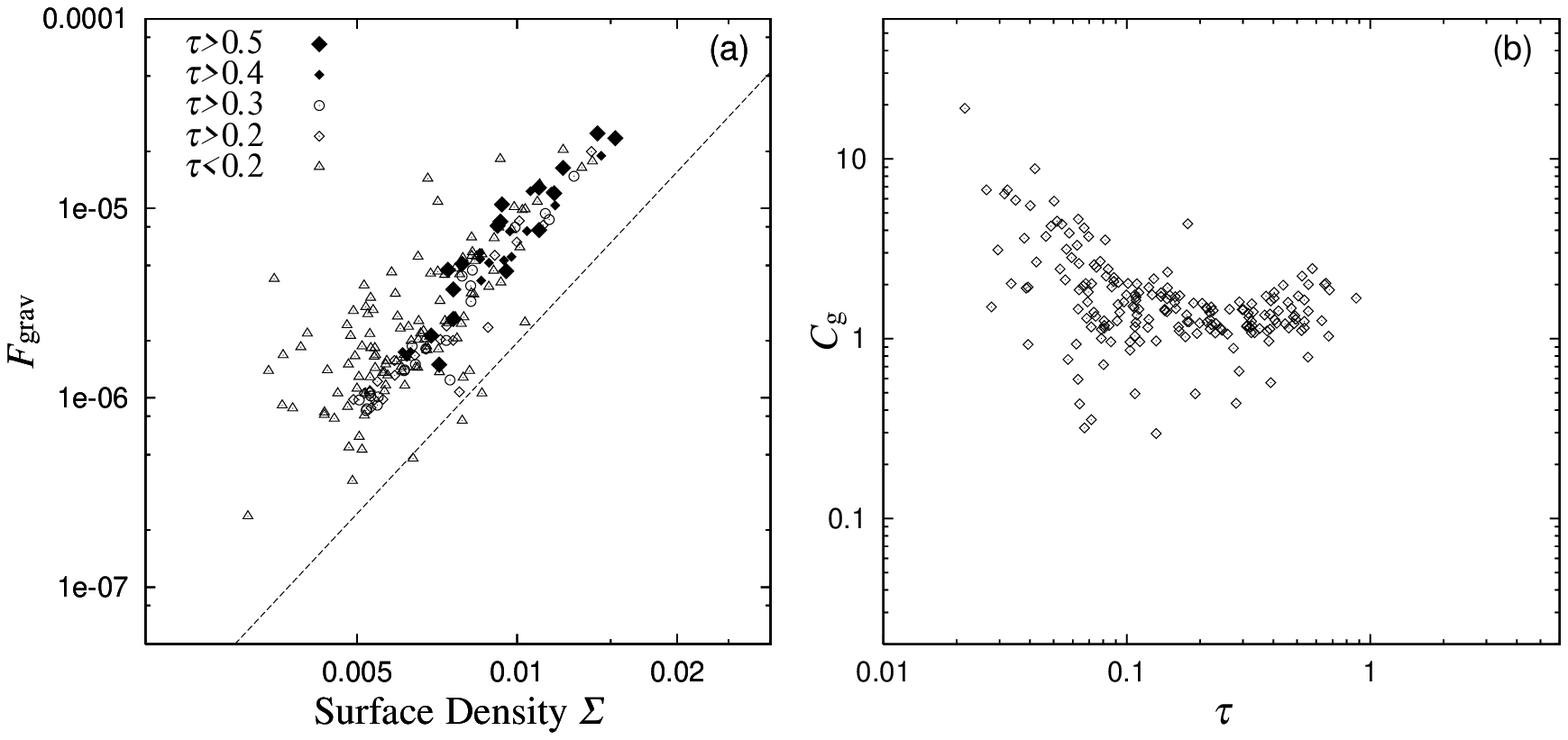}

Figs. \ref{Fgrav}
\end{figure}

\begin{figure}
\plotone{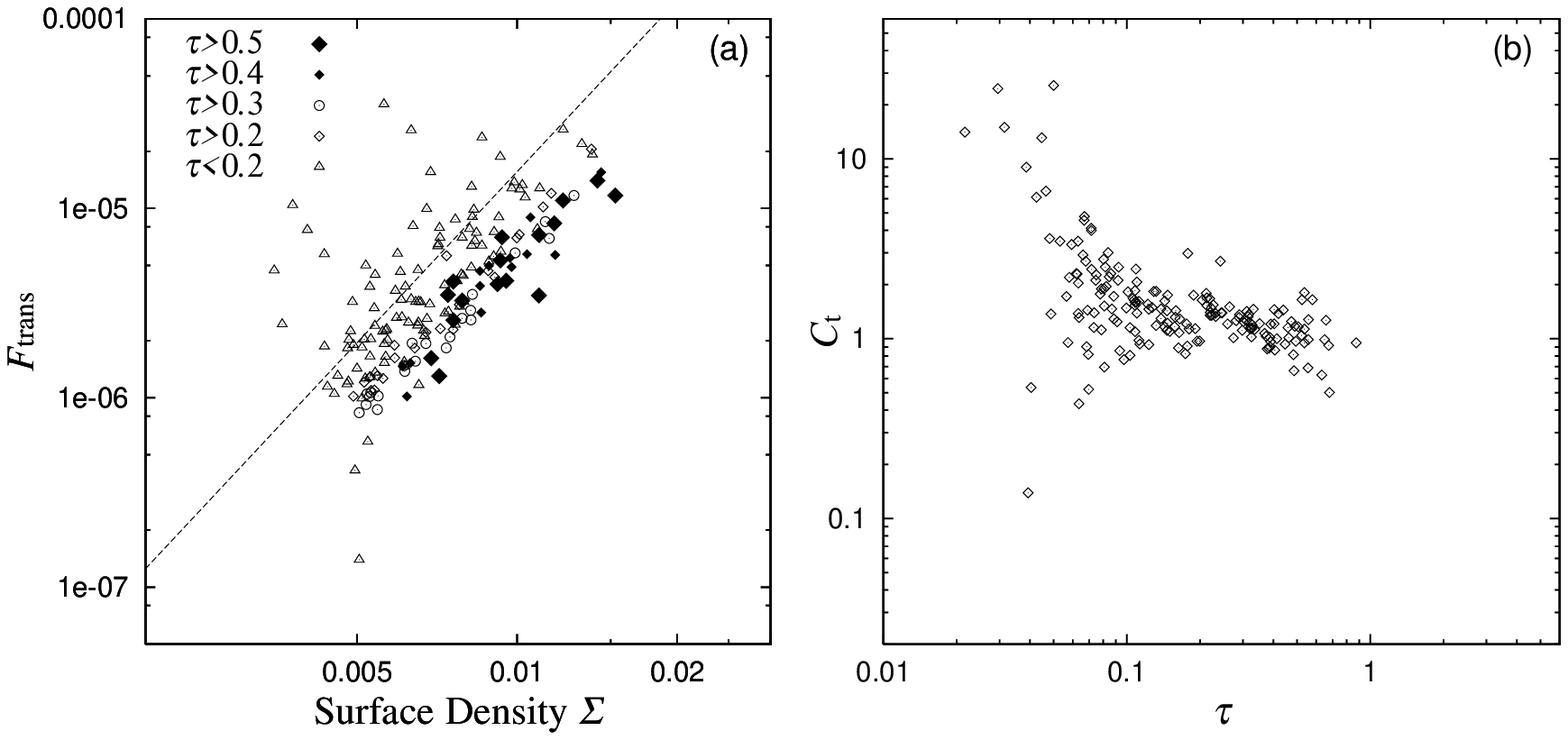}

Figs. \ref{Ftrans}
\end{figure}

\begin{figure}
\plotone{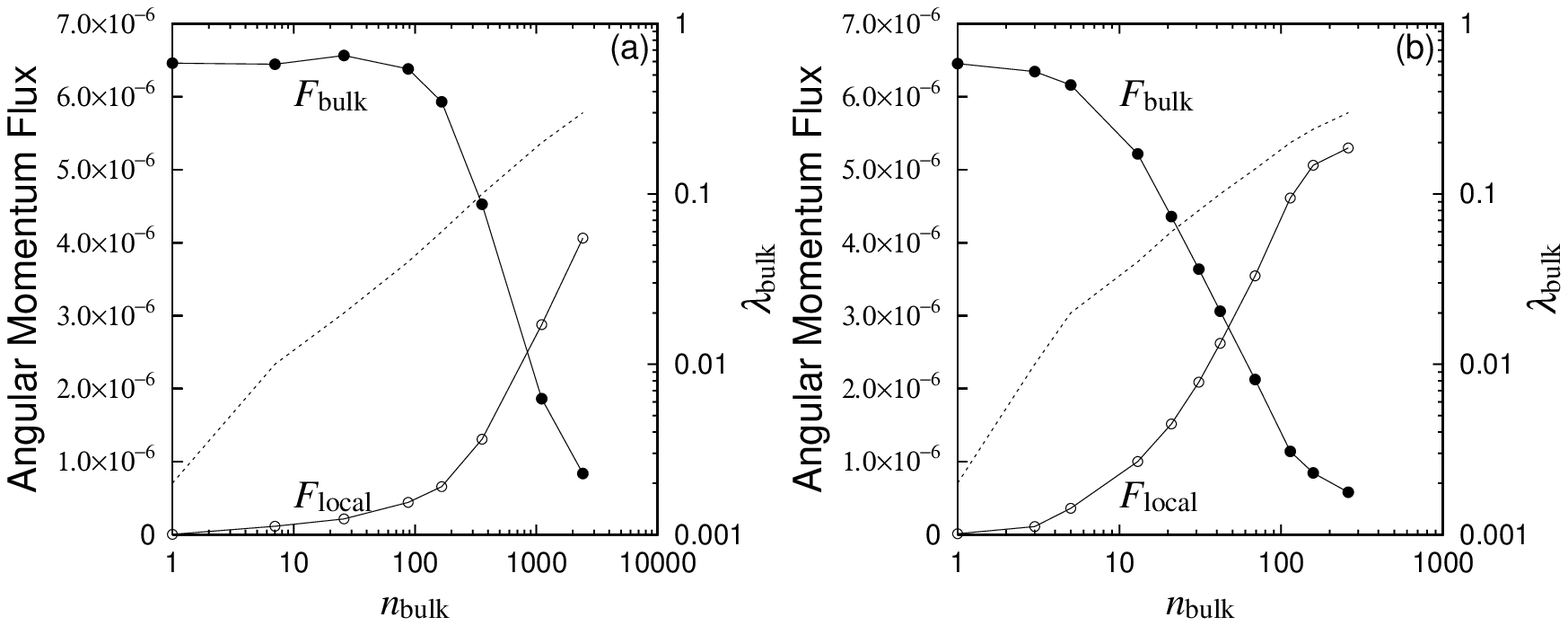}

Figs. \ref{FbulkandFlocal}
\end{figure}

\clearpage

\begin{figure}
\plottwo{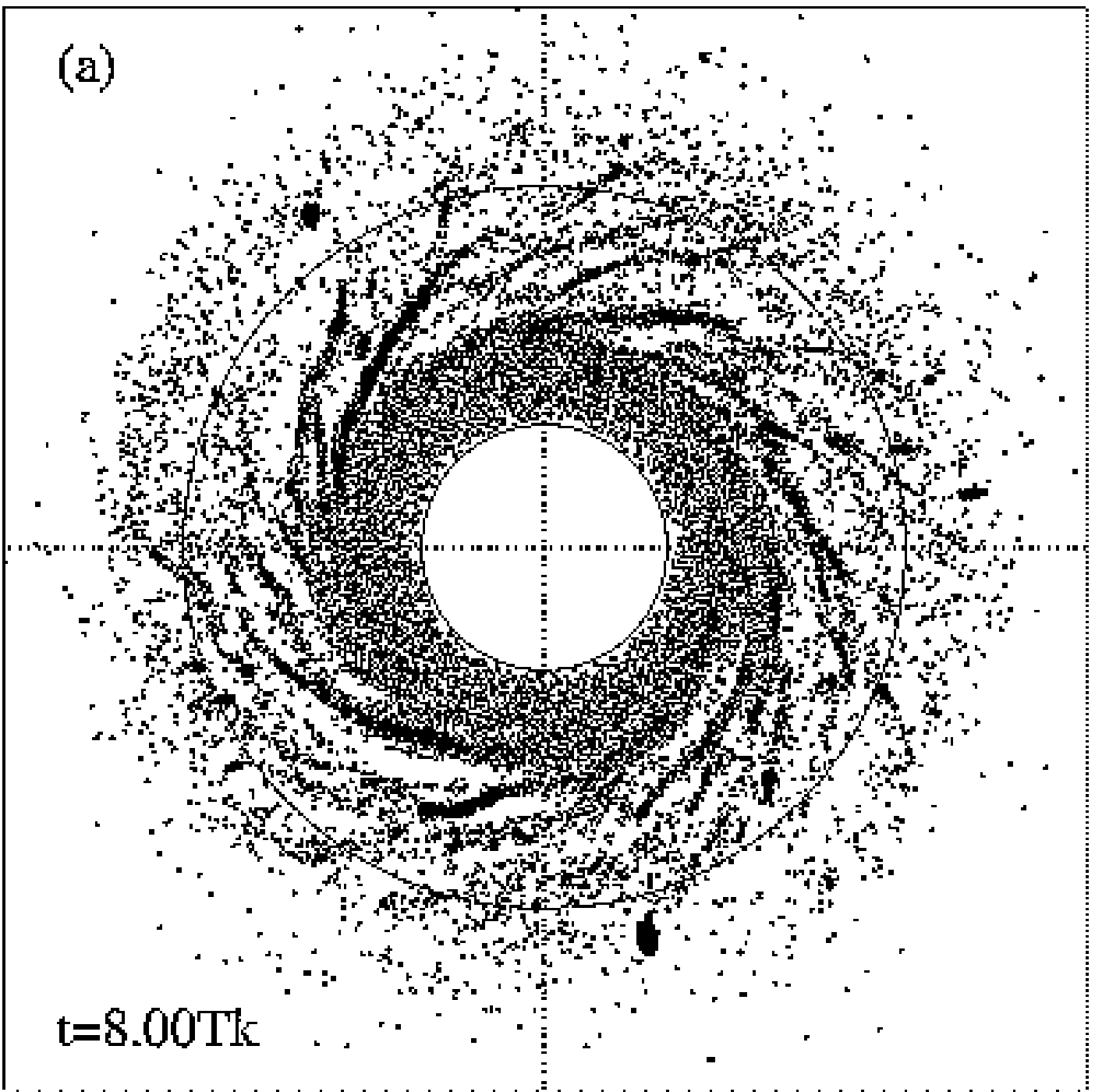}{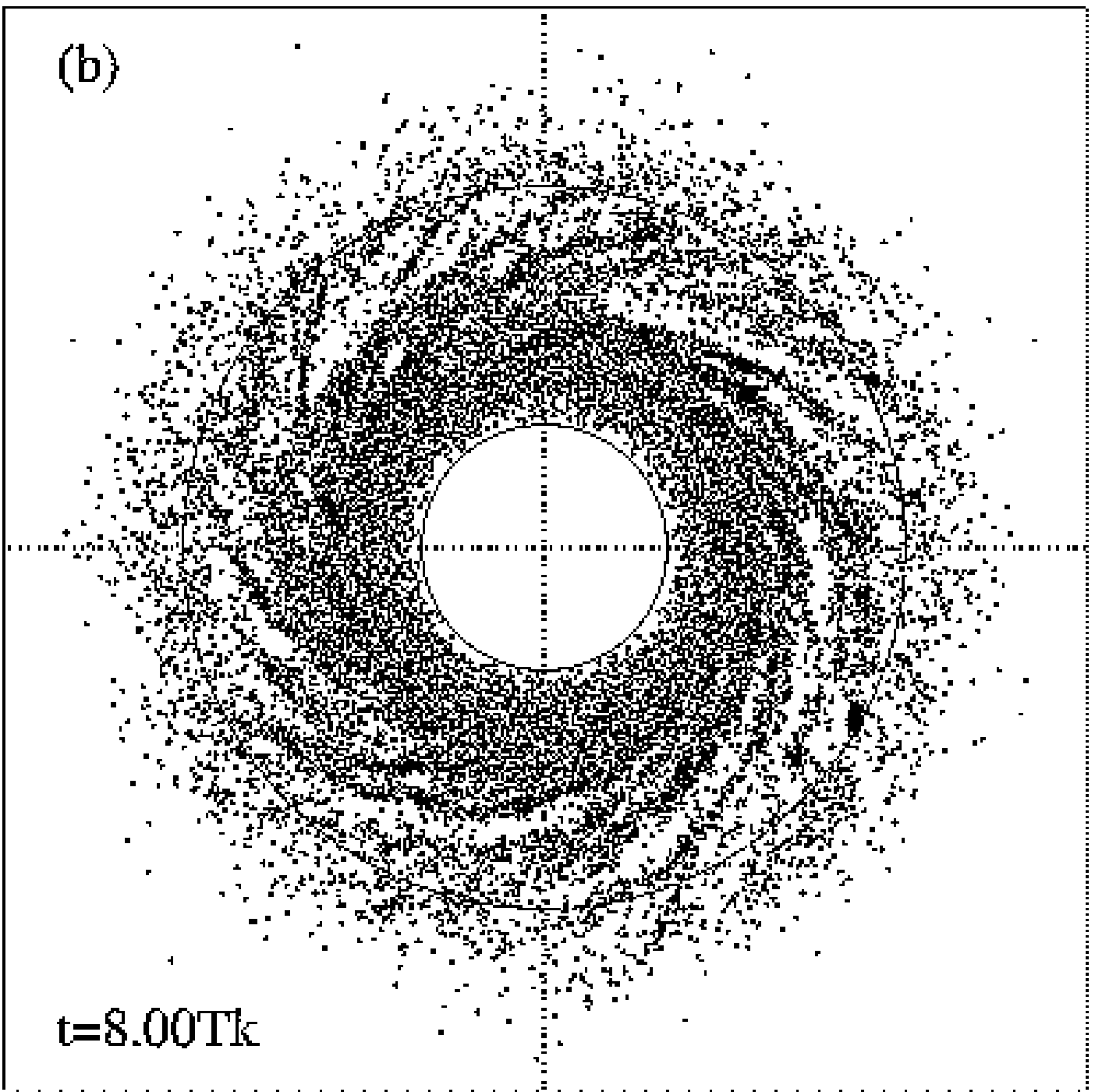}

Figs. \ref{snaprestitutions}
\end{figure}

\begin{figure}
\plotone{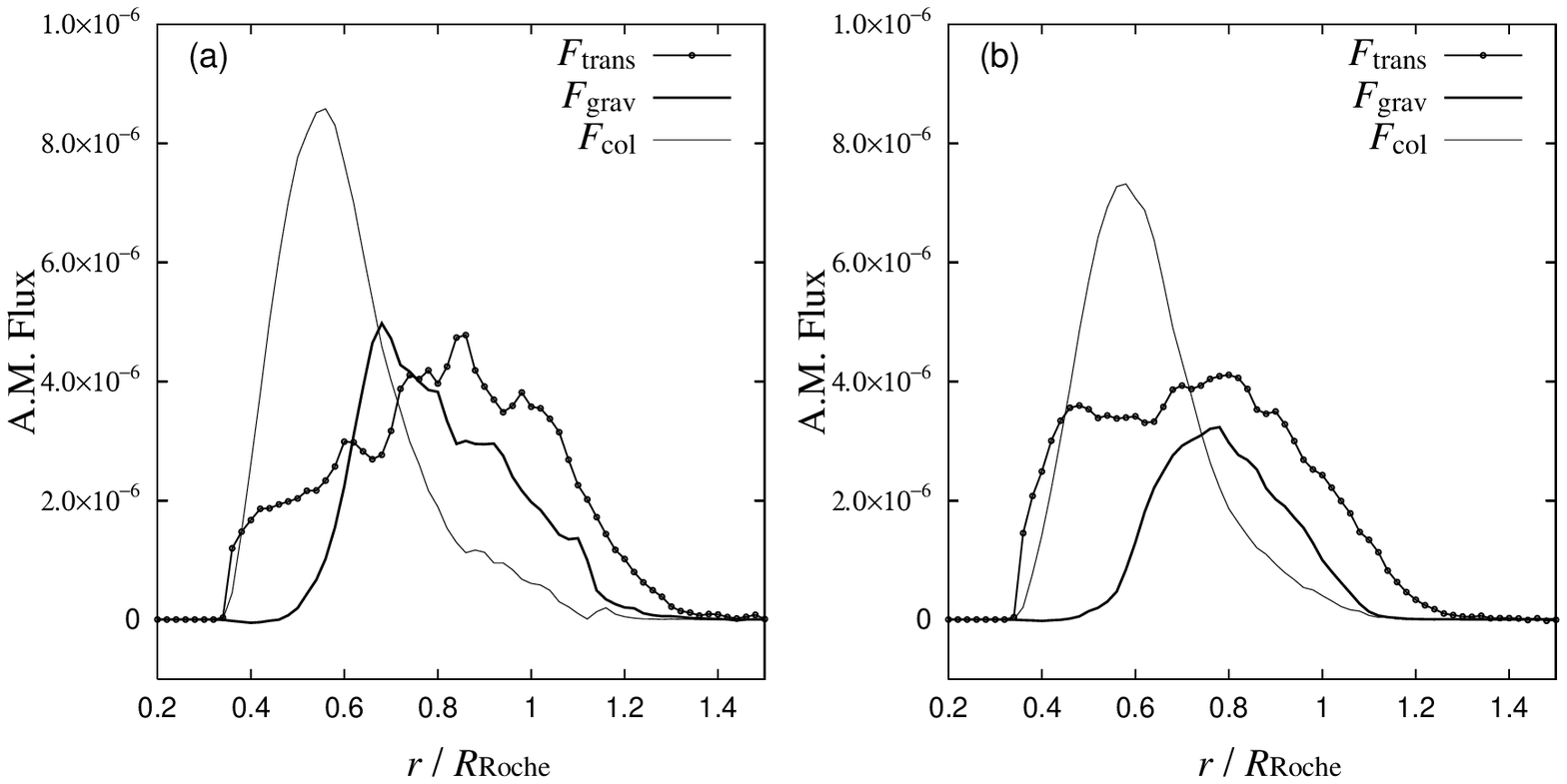}

Figs. \ref{fAMrestitution}
\end{figure}

\begin{figure}
\plotone{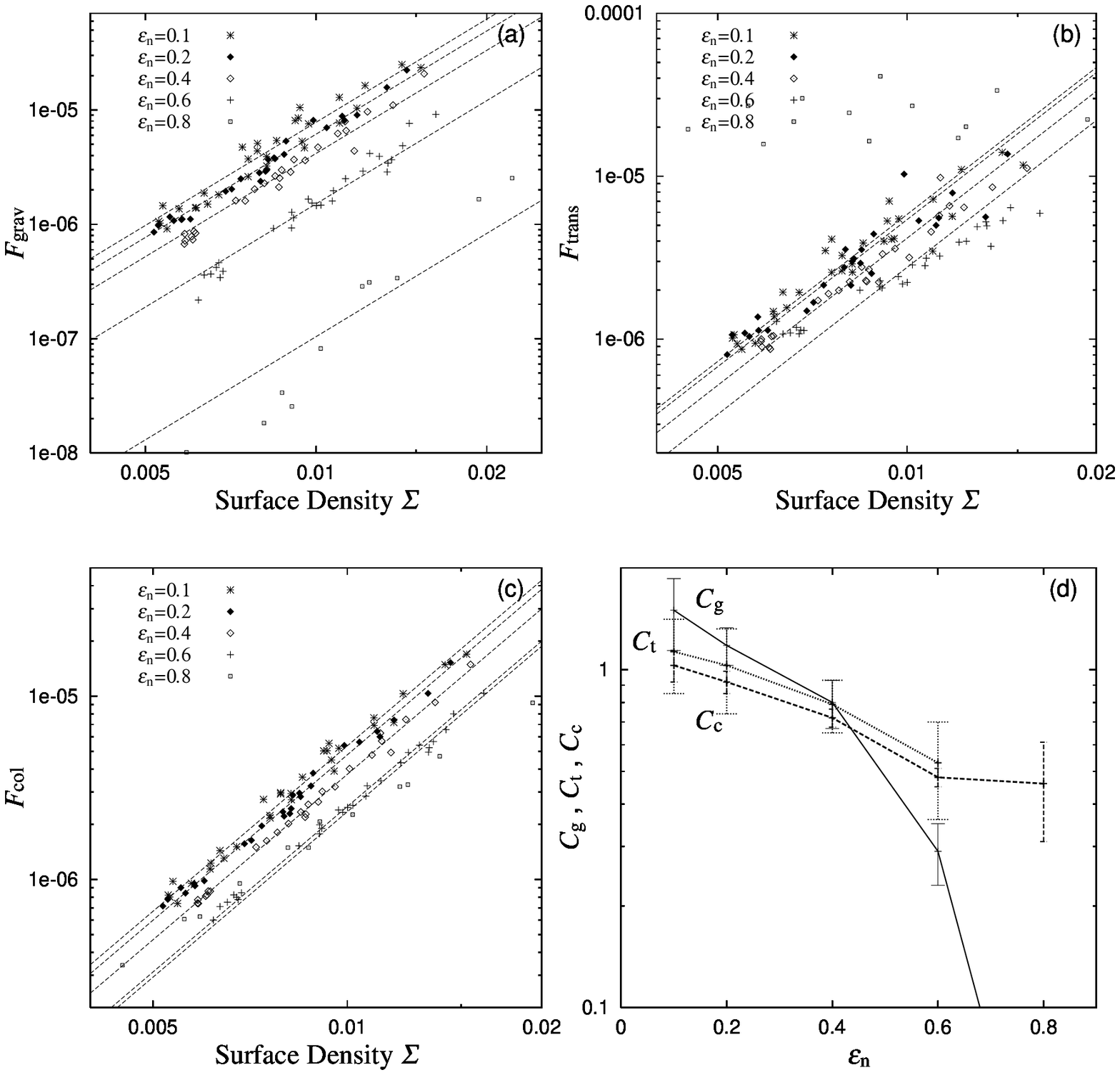}

Figs. \ref{Cgrestitution}
\end{figure}

\begin{figure}
\plotone{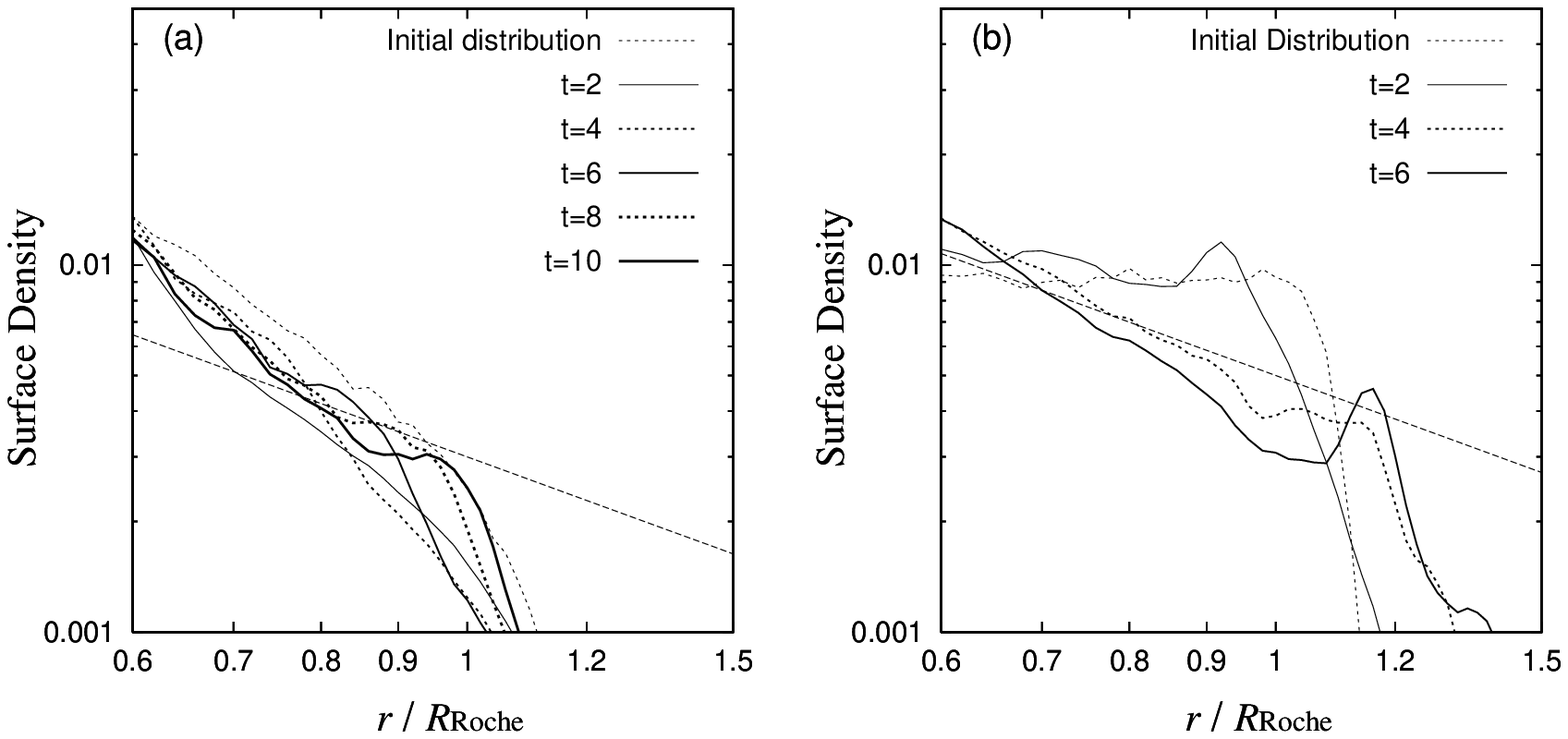}

Figs. \ref{surdenlog}
\end{figure}

\end{document}